\documentclass[fleqn,usenatbib]{mnras}
\usepackage{newtxtext,newtxmath,pifont}

\usepackage[T1]{fontenc}

\DeclareRobustCommand{\VAN}[3]{#2}
\let\VANthebibliography\thebibliography
\def\thebibliography{\DeclareRobustCommand{\VAN}[3]{##3}\VANthebibliography}

\newcommand{\gmax}{\mbox{$g$-band maximum}}

\newcommand{\tardis}{\protect \hbox {{\sc tardis}}}
\newcommand{\artis}{\protect \hbox {{\sc artis}}}
\newcommand{\CIIa}{C~{\sc ii} $\lambda 6580$}
\newcommand{\CIIb}{C~{\sc ii} $\lambda 7234$}
\newcommand{\SiIIa}{Si~{\sc ii} $\lambda 6355$}
\newcommand{\kyg}{SN 2020kyg}
\newcommand{\msun}{\mbox{M$_{\odot}$}}
\newcommand{\msol}{\mbox{M$_{\odot}$}}

\newcommand{\kms}{\mbox{$\rm{km}\,s^{-1}$}}

\newcommand{\zps}{\ensuremath{z_{\rm P1}}}

\newcommand{\wps}{\ensuremath{w_{\rm P1}}}
\newcommand{\grizy}{\ensuremath{grizy_{\rm P1}}}
\newcommand{\mch}{M$_{\rm Ch}$}

\usepackage{graphicx}	
\usepackage{amsmath}	

\title[Faint Iax supernova 2020kyg]{SN 2020kyg and 
the rates of faint Iax Supernovae from ATLAS}

\author[Srivastav et al.]{
Shubham Srivastav,$^{1}$\thanks{E-mail: S.Srivastav@qub.ac.uk}
S. J. Smartt,$^1$
M. E. Huber,$^2$
K. C. Chambers,$^2$
C. R. Angus,$^3$
T. -W. Chen,$^4$
\newauthor{
F. P. Callan,$^1$
J. H. Gillanders,$^1$
O. R. McBrien,$^5$
S. A. Sim,$^1$
M. Fulton,$^1$
J. Hjorth,$^3$
K. W. Smith,$^1$}
\newauthor{
D. R. Young,$^1$
K. Auchettl,$^{6,7,8}$
J. P. Anderson,$^{9}$
G. Pignata,$^{10,11}$
T.J.L. de Boer,$^2$
C.-C. Lin,$^2$}
\newauthor{
E. A. Magnier$^2$
}
\\ \\
$^{1}$Astrophysics Research Centre, School of Mathematics and Physics, Queen’s University Belfast, BT7 1NN, UK\\
$^2$Institute of Astronomy, University of Hawaii, 2680 Woodlawn Drive, Honolulu, HI 96822, USA\\
$^3$DARK, Niels Bohr Institute, University of Copenhagen, Lyngbyvej 2, DK-2100 Copenhagen \O, Denmark\\
$^4$The Oskar Klein Centre, Department of Astronomy, Stockholm University, AlbaNova, SE-10691 Stockholm, Sweden\\
$^5$Department of Physics, University of Warwick,
Coventry CV4 7AL, UK\\
$^6$School of Physics, The University of Melbourne, VIC 3010, Australia\\
$^7$ARC Centre of Excellence for All Sky Astrophysics in 3 Dimensions (ASTRO 3D)\\
$^8$Department of Astronomy and Astrophysics, University of California, Santa Cruz, CA 95064, USA\\
$^{9}$European Southern Observatory, Alonso de C\'{o}rdova 3107, Casilla 19, Santiago, Chile\\
$^{10}$Departamento de Ciencias Fisicas, Universidad Andres Bello, Fernandez Concha 700, Las Condes, Santiago, Chile\\
$^{11}$Millennium Institute of Astrophysics (MAS), Nuncio Monsenor
Sòtero Sanz 100, Providencia, Santiago, Chile
}

\date{Accepted XXX. Received YYY; in original form ZZZ}

\begin{document}
\graphicspath{{./}{figures/}}
\label{firstpage}
\pagerange{\pageref{firstpage}--\pageref{lastpage}}
\maketitle

\begin{abstract}

We present multi-wavelength follow-up observations of the ATLAS discovered faint Iax supernova SN 2020kyg that peaked at an absolute magnitude of $M_g \approx -14.9 \pm 0.2$, making it another member of the faint Iax supernova population. 
The bolometric light curve requires only $\approx 7 \times 10^{-3}$ \msol\ of radioactive $^{56}$Ni, with an ejected mass of $M_{\rm ej} \sim 0.4$ \msol\ and a low kinetic energy of $E \approx 0.05 \pm 0.02 \times 10^{51}$ erg.
We construct a homogeneous volume-limited sample of 902 transients observed by ATLAS within $100$ Mpc during a 3.5 year span. Using this sample, we constrain the rates of faint Iax ($M_r \gtrsim -16$) events within 60 Mpc at $12^{+14}_{-8}\%$ of the SN Ia rate. The overall Iax rate, at $15^{+17}_{-9}\%$ of the Ia rate, is dominated by the low-luminosity events, with luminous SNe Iax ($M_r \lesssim -17.5$) like 2002cx and 2005hk accounting for only $0.9^{+1.1}_{-0.5}\%$ of the Ia rate (a 2$\sigma$ upper limit of approximately 3\%). We favour the hybrid CONe WD + He star progenitor channel involving a failed deflagration of a near Chandrasekhar mass white dwarf, expected to leave a bound remnant and a surviving secondary companion, as a candidate explanation for faint Iax explosions. This scenario requires short delay times, consistent with the observed environments of SNe Iax. Furthermore, binary population synthesis calculations have suggested rates of $1-18\%$ of the SN Ia rate for this channel, consistent with our rate estimates.
\end{abstract}




\section{Introduction}

Supernovae of Type Ia (SNe Ia) are widely considered to be thermonuclear explosions involving at least one white dwarf (WD) star in a close binary system \citep{2014ARA&A..52..107M,2018PhR...736....1L}. SNe Ia are known for being remarkably homogeneous with standardisable light curves that enable them to be used as cosmic distance indicators \citep{1993ApJ...413L.105P,1996AJ....112.2391H}, leading to the discovery of the accelerating expansion of the universe \citep{1998AJ....116.1009R,1999ApJ...517..565P}. However, open questions regarding the nature of the progenitors and explosion mechanisms remain. Their use as cosmic probes notwithstanding, a rich diversity in the observed properties of SNe Ia is undeniably present \citep{2017hsn..book..317T}.

SNe Iax comprise a peculiar subclass of SNe Ia \citep{2003PASP..115..453L,2013ApJ...767...57F}, characterised by lower ejecta velocities of $\sim 2000-7000$ \kms\ \citep{2017hsn..book..375J}, as opposed to typical expansion velocity of $\gtrsim 10000$ \kms\ observed in normal SNe Ia around maximum \citep{2013ApJ...773...53F}. Despite the lower velocities, the early spectra of SNe Iax are known to show similarities with SNe Ia, in particular SN 1991T-like and SN 1999aa-like SNe Ia \citep{2013ApJ...767...57F,2014ApJ...786..134M}. Later spectra of SNe Iax however, unlike normal SNe Ia, tend to show persistent low-velocity permitted lines of Fe~{\sc ii} and Ca~{\sc ii} in addition to forbidden lines of [Fe~{\sc ii}], [Ni~{\sc ii}] and [Ca~{\sc ii}] \citep[eg.][]{2006AJ....132..189J,2008ApJ...680..580S}. SNe Iax constitute a very heterogeneous subclass with rich diversity in line velocities, relative strengths and line widths of permitted and forbidden transitions, especially in later spectra \citep{2014ApJ...786..134M,2015A&A...573A...2S,2015ApJ...806..191Y,2016MNRAS.461..433F}. \citet{2017A&A...601A..62M} noted that spectra of the Iax event PS1-12bwh around maximum closely resembled spectra of SN 2005hk at phases approximately a week earlier, implying lower densities for the high velocity ejecta in PS1-12bwh relative to SN 2005hk.
This diversity is also reflected in photometric properties, with SNe Iax exhibiting a wide range of decline rates, rise times and peak luminosities \citep{2016A&A...589A..89M}. 

The nearby Iax SN 2012Z is the only SN Ia with an identified progenitor system in deep pre-explosion HST images, where the blue source was identified as a helium-rich companion to the WD \citep{2014Natur.512...54M}. 
Follow-up HST imaging $\sim1400$ days after explosion shows that the source has not disappeared \citep{2021arXiv210604602M}. A potential companion was also detected in the case of SN 2008ha by \citet{2014ApJ...792...29F} in HST images $\sim 4$ years after explosion.
Pure deflagration models \citep{2012ApJ...761L..23J,2013MNRAS.429.2287K} of Chandrasekhar mass (\mch) WDs in a single degenerate system that result in either partial or complete disruption of the WD have had success in reproducing the low kinetic energies and range of luminosity (except for the faint end of the distribution) observed in SNe Iax. These weak `failed' explosions are thus thought to leave behind a bound remnant enriched with the deflagration ashes \citep{2012ApJ...761L..23J,2014MNRAS.438.1762F}. A similar model involving a hybrid carbon-oxygen-neon (CONe) WD instead of a CO WD was invoked by \citet{2015MNRAS.450.3045K} to explain the faintest members of the subclass. Recent simulations by \citet{2021arXiv210902926L} show that deflagrations of CO WDs can also produce faint explosions with peak absolute magnitudes of $M_r \approx -15$. The bound remnant picture is promising, since an optically thick wind from the remnant \citep{2016MNRAS.461..433F,2017ApJ...834..180S} would help explain the long-lived photosphere and the fact that the spectra don't become nebular even at later epochs. It can also potentially explain the late-time flattening observed in the light curves of SNe Iax  \citep[eg.][]{2014ApJ...786..134M,2018MNRAS.474.2551S,2018PASJ...70..111K,2021PASJ...73.1295K,2021arXiv210604602M}. However, the radiation physics involving the bound remnant is complicated and our current theoretical insights regarding the long term evolution of these bound remnants, or postgenitors, are limited \citep{2019ApJ...872...29Z}.

Of particular interest are the least luminous and least energetic members of the subclass -- events like SNe 2021fcg \citep{2021ApJ...921L...6K}, 2008ha \citep{2009AJ....138..376F,2009Natur.459..674V,2014A&A...561A.146S}, 2010ae \citep{2014A&A...561A.146S} and 2019gsc \citep{2020ApJ...892L..24S,2020MNRAS.496.1132T}. 
With $M_g \lesssim -14$, SN 2019gsc synthesized a meagre $2 \times 10^{-3}$ \msol\ of radioactive $^{56}$Ni. These `faint Iax' events show peak luminosity and kinetic energy that is lower than that of normal SNe Ia by a factor of a few hundred, posing a challenge for theoretical models to reproduce these extreme properties. Although a core-collapse origin for SN 2008ha was proposed by \citet{2009Natur.459..674V}, the photometric and spectroscopic properties of these faint objects seem to align more closely with the overall population of SNe Iax \citep{2017hsn..book..375J}. In particular, the unambiguous detection of Co~{\sc ii} lines in the Near Infrared (NIR) spectra of lower luminosity Iax events SN 2010ae \citep{2014A&A...561A.146S} and SN 2019muj \citep{2021MNRAS.501.1078B}, also seen in more luminous members like SN 2014ck \citep{2016MNRAS.459.1018T}, suggests a kinship with more luminous SNe Iax and the general Ia population. 
Although the deflagration model of a hybrid CONe WD  \citep{2015MNRAS.450.3045K} yields extremely faint transients with peak $B$-band luminosity between $-13.2$ and $-14.6$, consistent with faint SNe Iax, it also produces very low ejecta masses ($\sim 0.01$ \msol), an order of magnitude lower than that inferred from observations, leading to much faster predicted decline rate.

The volumetric rates of SNe Iax are not particularly well-constrained, ranging from roughly 5 to 30 percent of the total SN Ia rate \citep{2011MNRAS.412.1441L,2013ApJ...767...57F}. Also, SNe Iax seem to occur preferentially in young stellar populations \citep[eg.][]{2018MNRAS.473.1359L,2020MNRAS.499.1424H}, suggesting the progenitors have short delay times \citep{2020MNRAS.493..986T}, in turn suggesting more massive WD progenitors, that would attain \mch\ in a shorter time \citep{2017hsn..book..375J}. An intriguing alternative to the single degenerate scenario for SNe Iax is a double degenerate merger scenario involving a primary ONe WD and a secondary CO WD \citep{2018ApJ...869..140K}; or a WD and a neutron star or black hole \citep[NS/BH,][]{2013ApJ...763..108F}. The ONe--CO WD merger model of \citet{2018ApJ...869..140K} yields an extremely faint transient with $M_V = -11.3$, roughly three times less luminous than SN 2021fcg, the faintest Iax event discovered to date \citep{2021ApJ...921L...6K}.
A ONe WD - NS/BH merger explored by \citet{2021arXiv210403415B} can theoretically  produce relatively faint (up to $M \sim -16.5$), but rather long-lived and red transients. There is one observed example of such a proposed merger, AT 2018kzr \citep{2019ApJ...885L..23M,2020MNRAS.497..246G}. However this transient was fast declining and relatively blue \citep{2019ApJ...885L..23M}, and the composition from spectral modelling suggested a significant amount of $^{54}$Fe was present \citep{2020MNRAS.497..246G}. Hence there is no clear consensus on what  a WD + NS/BH merger should look like in the optical and NIR. 

In this paper, we present results of multi-wavelength observations of the faint Iax \kyg. We also constrain the volumetric rates of Iax and faint Iax events using a 3.5 year sample of transients within 100 Mpc observed by the ATLAS survey.

%

\section{Discovery and Follow-up Observations}

We discovered \kyg\ (as ATLAS20nuc)  with the Asteroid Terrestrial-impact Last Alert System   \citep[ATLAS;][]{2018PASP..130f4505T} 
in images taken on 2020 May 24.4 UT 
or MJD 58993.4, at an orange-band magnitude of $o=18.83$ \citep{2020TNSAN.113....1S}. The data from ATLAS are 
processed in real time, initially on site and then the detections are filtered on our ATLAS Transient Server at Queen's University Belfast \citep{2020PASP..132h5002S}.  \kyg\ was discovered within the same night (as is now routine), 
with a human scanner recognising it 3.6\,hrs after the quad of nightly ATLAS images at that position was completed.
We registered the discovery on the IAU Transient Name Server as AT 2020kyg on 2020 May 24.55 UT, noting that it was a young 
object likely associated with host galaxy NGC 5012, with a non-detection in ATLAS 4 days before \citep{2020TNSAN.113....1S}. 
It was rapidly classified as a SN Iax independently by \citet{2020ATel13761....1O} and \citet{2020TNSCR1559....1H}
with spectra taken 0.40 and 0.88 days after the discovery announcement on the TNS, respectively. ATLAS continued to observe the field until MJD 59077 (2020 August 16), or 84 days after discovery, providing good photometric sampling. We supplemented this with multi-colour optical to NIR photometry from different observing facilities discussed below. The ATLAS data are publicly available from our forced photometry server\footnote{https://fallingstar.com/forcedphot} \citep{2021TNSAN...7....1S}.

\subsection{Photometry}
Following the classification, we triggered initial follow-up photometry from the Robotic 2m Liverpool Telescope \citep{2004SPIE.5489..679S}.
Images were obtained in the Sloan $ugriz$ bands using the 4k x 4k IO:O instrument. Basic data reductions, including bias subtraction, overscan trimming and flat-field corrections were performed automatically by the IO:O pipeline and the processed images were downloaded from the LT data archive. Point-spread function (PSF) photometry was performed on the images and the SN magnitudes were estimated by calibrating the zero-points against PS1 reference stars in the field of view. No image subtraction was applied since the SN was relatively bright at this epoch and is significantly offset ($\sim 35 \arcsec$) from it's host.

As the transient evolved and faded, optical photometric follow-up was obtained using the 1.8m Pan-STARRS1 (PS1) telescope \citep{Chambers2016}. PS1 uses the 1.4 Gigapixel camera GPC1 with a pixel scale of $0.26\arcsec$. Photometry was obtained in the \grizy\ filter system described in \citet{2012ApJ...750...99T}. The images were processed with the Image Processing Pipeline \citep[IPP;][]{2020ApJS..251....3M} and image subtraction was performed using the PS1 Science Consortium \citep[PS1SC;][]{Chambers2016} 3$\pi$ survey data as reference. Instrumental magnitudes were computed using PSF photometry and calibrations were performed using zero-points calculated using PS1 reference stars in the field \citep{2020ApJS..251....4W,2020ApJS..251....5M}. 
We also triggered NIR photometry in $JHK$ bands on the Wide Field Infrared Camera (WFCAM) on the United Kingdom Infrared Telescope (UKIRT) at Maunakea, Hawaii. WFCAM has four 2048$\times$2048 HgCdTe detectors, with a pixel scale of $0.4\arcsec$ and a 0.2 square degree field of view. The processed data was obtained from the Cambridge Astronomy Survey Unit (CASU). The dithered frames in each filter were resampled and co-added using the \texttt{SWarp} package \citep{2002ASPC..281..228B}. PSF photometry was performed on the co-added frames and the zero-points were calibrated using the 2MASS catalog \citep{2006AJ....131.1163S}. No image subtraction was done for the NIR images. 

\kyg\ was followed up by the UVOT instrument \citep{2005SSRv..120...95R} aboard the \textit{Swift} observatory \citep{2004ApJ...611.1005G}. The images were obtained in broadband filters $v$ (5468 \AA), $b$ (4392 \AA), $u$ (3465 \AA), $uvw1$ (2600 \AA), $uvm2$ (2246 \AA) and $uvw2$ (1928 \AA). The processed images were downloaded from the \textit{Swift} archive. The individual frames for each observation ID in a given filter were co-added using the \texttt{uvotimsum} task within the High Energy Astrophysics SOFTware (\textsc{heasoft}) package. Photometry was performed on the co-added frames with the \texttt{uvotsource} task using a $5\arcsec$ aperture, following the recipe of \citet{2008MNRAS.383..627P} and \citet{2009AJ....137.4517B}, using updated zero-points and effective area curves for the UVOT filters \citep{doi:10.1063/1.3621807}. Since archival images of the field were available in $u$, $uvw1$, $uvm2$ and $uvw2$, the magnitudes in these filters were estimated after subtraction of the underlying galaxy flux. To perform the template subtraction, we adopt the method outlined by \citet{2014Ap&SS.354...89B}. The count rates in the template frames (for a $5\arcsec$ aperture) were subtracted from the count rates in the SN frames, before applying an aperture correction. The detections in the $uvw1$, $uvm2$ and $uvw2$ images are marginal ($<5\sigma$), and we only use the $u$ magnitudes in the subsequent analysis.
Once the supernova had faded significantly in the $u$-band ($u_{\rm AB} > 21$), we triggered deep $u$-band imaging observations on the 3.6m Canada-France-Hawaii Telescope (CFHT) at Maunakea. The images were obtained with the wide-field MegaCam instrument consisting of 40 2048$\times$4612 pixel CCDs with a $0.2\arcsec$ pixel scale and a 1 square degree field of view. The processed and astrometrically calibrated images were obtained from the Data Archiving and Distribution System (DADS). For each epoch, the individual exposures were aligned and co-added using \texttt{SWarp} and PSF photometry was performed to estimate instrumental magnitudes. The SN instrumental magnitudes were calibrated using zero-points calculated from local reference stars in the SDSS catalog. The CFHT images are much deeper than the SDSS $u$-band reference frames of the field, thus no image subtraction was performed. 
The photometric magnitudes are summarised in Table~\ref{tab:photometry}. 

\begin{table*}
\caption{Summary of photometric observations in $ugrizyJHK$ bands for \kyg. All magnitudes are in the AB system.}
\resizebox{\textwidth}{!}{\begin{tabular}{ccccccccccc}\hline
MJD & $u$ & $g$ & $r$ & $i$ & $z$ & $y$ & $J$ & $H$ & $K$ & Instrument \\
\hline
58991.26 & $-$ & $20.03\pm0.35$ & $20.03\pm0.14$ & $-$ & $-$ & $-$ & $-$ & $-$ & $-$ & ZTF\\
58994.94 & $-$ & $18.43\pm0.01$ & $18.49\pm0.02$ & $18.66\pm0.03$ & $18.82\pm0.028$ & $-$ & $-$ & $-$ & $-$ & LT\\
58995.22 & $18.86\pm0.12$ & $-$ & $-$ & $-$ & $-$ & $-$ & $-$ & $-$ & $-$ & UVOT\\
58996.03 & $18.96\pm0.12$ & $-$ & $-$ & $-$ & $-$ & $-$ & $-$ & $-$ & $-$ & UVOT\\
58998.75 & $19.36\pm0.17$ & $-$ & $-$ & $-$ & $-$ & $-$ & $-$ & $-$ & $-$ & UVOT\\
58999.28 & $19.59\pm0.16$ & $-$ & $-$ & $-$ & $-$ & $-$ & $-$ & $-$ & $-$ & UVOT\\
58999.89 & $-$ & $18.39\pm0.02$ & $18.29\pm0.02$ & $18.40\pm0.03$ & $18.46\pm0.03$ & $-$ & $-$ & $-$ & $-$ & LT\\
59000.89 & $-$ & $18.50\pm0.02$ & $18.32\pm0.02$ & $18.40\pm0.03$ & $18.43\pm0.03$ & $-$ & $-$ & $-$ & $-$ & LT\\
59003.39 & $20.30\pm0.20$ & $-$ & $-$ & $-$ & $-$ & $-$ & $-$ & $-$ & $-$ & UVOT\\
59005.18 & $21.31\pm0.27$ & $-$ & $-$ & $-$ & $-$ & $-$ & $-$ & $-$ & $-$ & UVOT\\
59006.34 & $-$ & $19.05\pm0.07$ & $18.36\pm0.03$ & $18.42\pm0.03$ & $18.35\pm0.02$ & $18.48\pm0.05$ & $-$ & $-$ & $-$ & PS1\\
59009.11 & $21.65\pm0.28$ & $-$ & $-$ & $-$ & $-$ & $-$ & $-$ & $-$ & $-$ & UVOT\\
59010.24 & $-$ & $-$ & $-$ & $-$ & $-$ & $-$ & $19.36\pm0.08$ & $19.40\pm0.18$ & $19.86\pm0.33$ & UKIRT\\
59010.36 & $-$ & $19.65\pm0.09$ & $18.63\pm0.03$ & $18.61\pm0.03$ & $18.56\pm0.06$ & $18.57\pm0.11$ & $-$ & $-$ & $-$ & PS1\\
59013.34 & $-$ & $20.00\pm0.08$ & $18.77\pm0.04$ & $18.80\pm0.05$ & $18.76\pm0.09$ & $18.74\pm0.23$ & $-$ & $-$ & $-$ & PS1\\
59014.27 & $-$ & $-$ & $-$ & $-$ & $-$ & $-$ & $19.40\pm0.18$ & $19.53\pm0.18$ & $-$ & UKIRT\\
59016.34 & $-$ & $20.24\pm0.07$ & $19.02\pm0.03$ & $18.88\pm0.03$ & $18.87\pm0.05$ & $19.02\pm0.11$ & $-$ & $-$ & $-$ & PS1\\
59017.31 & $-$ & $-$ & $-$ & $-$ & $-$ & $-$ & $19.49\pm0.08$ & $-$ & $20.07\pm0.31$ & UKIRT\\
59020.27 & $22.70\pm0.18$ & $-$ & $-$ & $-$ & $-$ & $-$ & $-$ & $-$ & $-$ & CFHT\\
59021.28 & $-$ & $20.40\pm0.05$ & $19.26\pm0.02$ & $19.15\pm0.02$ & $19.01\pm0.03$ & $19.12\pm0.07$ & $-$ & $-$ & $-$ & PS1\\
59021.29 & $22.82\pm0.17$ & $-$ & $-$ & $-$ & $-$ & $-$ & $-$ & $-$ & $-$ & CFHT\\
59024.24 & $-$ & $-$ & $-$ & $-$ & $-$ & $-$ & $19.76\pm0.14$ & $19.96\pm0.19$ & $-$ & UKIRT\\
59024.33 & $23.05\pm0.19$ & $-$ & $-$ & $-$ & $-$ & $-$ & $-$ & $-$ & $-$ & CFHT\\
59025.26 & $-$ & $-$ & $-$ & $-$ & $-$ & $-$ & $19.88\pm0.11$ & $19.96\pm0.19$ & $-$ & UKIRT\\
59026.27 & $-$ & $20.71\pm0.10$ & $19.58\pm0.04$ & $19.48\pm0.05$ & $19.27\pm0.06$ & $19.41\pm0.13$ & $-$ & $-$ & $-$ & PS1\\
59029.27 & $-$ & $20.86\pm0.09$ & $19.66\pm0.04$ & $19.57\pm0.03$ & $19.21\pm0.03$ & $19.52\pm0.09$ & $-$ & $-$ & $-$ & PS1\\
59032.30 & $-$ & $-$ & $19.85\pm0.08$ & $19.74\pm0.09$ & $19.43\pm0.09$ & $19.44\pm0.18$ & $-$ & $-$ & $-$ & PS1\\
59034.27 & $-$ & $-$ & $-$ & $-$ & $-$ & $-$ & $20.41\pm0.15$ & $-$ & $-$ & UKIRT\\
59035.36 & $-$ & $-$ & $20.01\pm0.30$ & $19.74\pm0.16$ & $19.42\pm0.15$ & $-$ & $-$ & $-$ & $-$ & PS1\\
59036.27 & $-$ & $-$ & $-$ & $-$ & $-$ & $-$ & $-$ & $20.35\pm0.23$ & $-$ & UKIRT\\
59037.29 & $-$ & $-$ & $-$ & $-$ & $-$ & $-$ & $-$ & $-$ & $21.03\pm0.35$ & UKIRT\\
59040.28 & $-$ & $21.08\pm0.16$ & $20.12\pm0.14$ & $19.94\pm0.13$ & $19.80\pm0.18$ & $-$ & $-$ & $-$ & $-$ & PS1\\
59041.26 & $-$ & $-$ & $-$ & $-$ & $-$ & $-$ & $20.60\pm0.14$ & $-$ & $-$ & UKIRT\\
59041.28 & $23.69\pm0.23$ & $-$ & $-$ & $-$ & $-$ & $-$ & $-$ & $-$ & $-$ & CFHT\\
59042.26 & $23.92\pm0.30$ & $-$ & $-$ & $-$ & $-$ & $-$ & $-$ & $-$ & $-$ & CFHT\\
59043.27 & $23.85\pm0.23$ & $-$ & $-$ & $-$ & $-$ & $-$ & $-$ & $-$ & $-$ & CFHT\\
59047.29 & $-$ & $21.22\pm0.10$ & $20.21\pm0.06$ & $20.09\pm0.09$ & $19.71\pm0.10$ & $20.06\pm0.26$ & $-$ & $-$ & $-$ & PS1\\
59048.26 & $-$ & $-$ & $-$ & $-$ & $-$ & $-$ & $20.81\pm0.19$ & $-$ & $-$ & UKIRT\\
59048.29 & $24.02\pm0.22$ & $-$ & $-$ & $-$ & $-$ & $-$ & $-$ & $-$ & $-$ & CFHT\\
59051.26 & $24.13\pm0.32$ & $-$ & $-$ & $-$ & $-$ & $-$ & $-$ & $-$ & $-$ & CFHT\\
59052.28 & $-$ & $21.37\pm0.26$ & $20.39\pm0.13$ & $20.36\pm0.13$ & $19.83\pm0.14$ & $19.82\pm0.30$ & $-$ & $-$ & $-$ & PS1\\
59054.26 & $-$ & $-$ & $-$ & $-$ & $-$ & $-$ & $-$ & $20.84\pm0.23$ & $-$ & UKIRT\\
59058.28 & $-$ & $-$ & $20.44\pm0.26$ & $20.48\pm0.20$ & $-$ & $-$ & $-$ & $-$ & $-$ & PS1\\
59064.28 & $-$ & $-$ & $20.53\pm0.32$ & $20.53\pm0.26$ & $20.30\pm0.28$ & $-$ & $-$ & $-$ & $-$ & PS1\\
59074.26 & $-$ & $-$ & $20.78\pm0.23$ & $20.64\pm0.19$ & $20.34\pm0.25$ & $-$ & $-$ & $-$ & $-$ & PS1\\
\hline
\end{tabular}}
\label{tab:photometry}
\end{table*}

\begin{figure*}
	\includegraphics[width=0.85\linewidth]{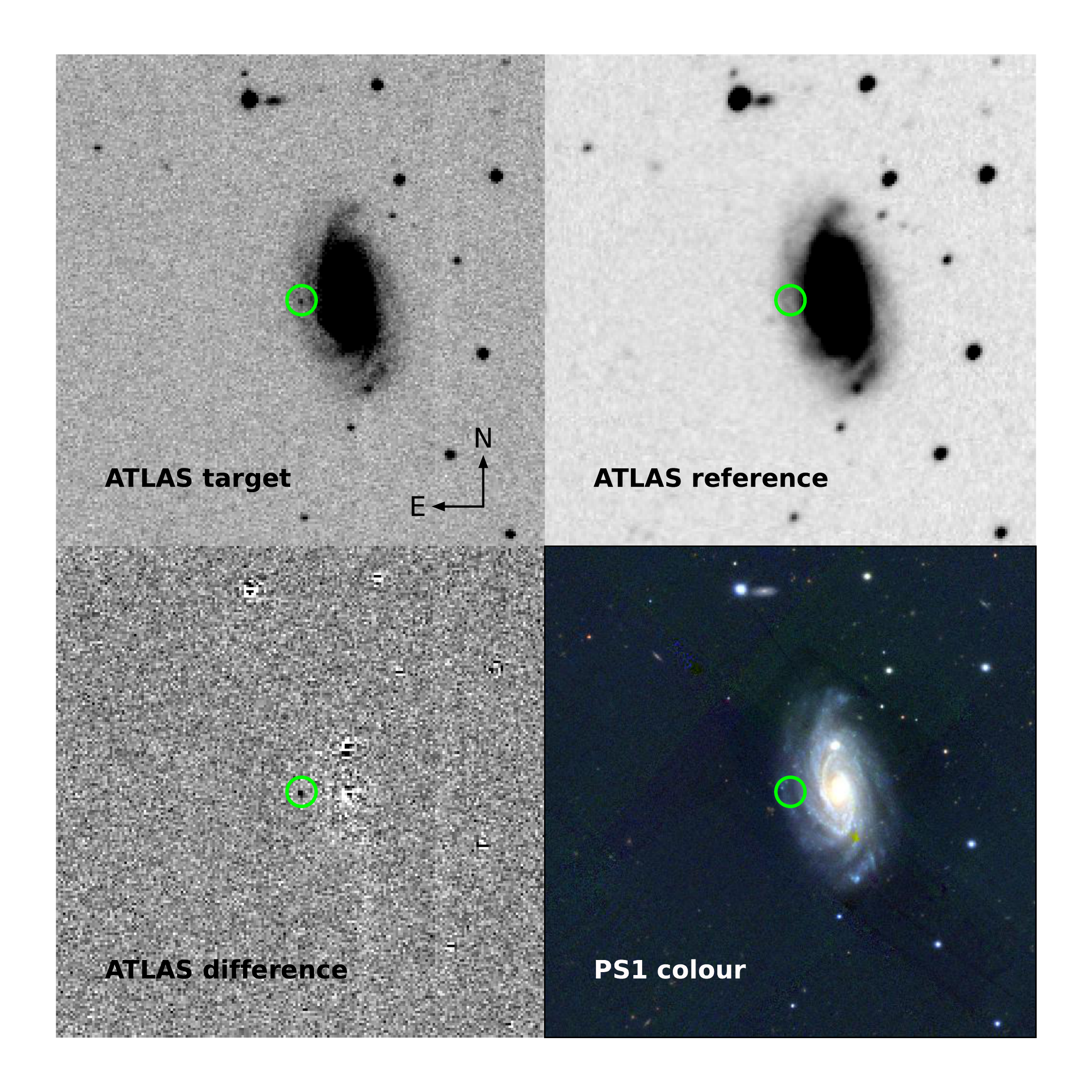}
    \caption{ATLAS target, reference and difference images for \kyg. Also shown is the color-composite PS1 image.}
    \label{fig:image}
\end{figure*}

\subsection{Spectroscopy}

The initial two epochs of spectroscopic observations  on 2020 May 30 and May 31 were obtained with the SPectrograph for the Rapid Acquisition of Transients \citep[SPRAT;][]{2014SPIE.9147E..8HP} on the LT while the SN was relatively bright. The wavelength and flux calibrated 1D spectra were obtained from the SPRAT pipeline. The default extraction parameters on the pipeline generally work well for isolated point sources. Since \kyg\ is significantly offset from its host galaxy, the quality of the extraction was adequate and a manual extraction in \textsc{iraf}\footnote{IRAF is distributed by the National Optical Astronomy Observatory, which is operated by the Association of Universities for Research in Astronomy (AURA) under a cooperative agreement with the National Science Foundation.} does not discernibly enhance the signal-to-noise. 

Subsequent spectroscopic observations during two epochs were obtained with the Alhambra Faint Object Spectrograph and Camera (ALFOSC) on the Nordic Optical Telescope (NOT) at La Palma on 2020 June 13 and June 21 using grism 4 and a 1.0\arcsec\ slit, aligned along the parallactic angle. Both observations were conducted under clear observing conditions and good seeing. The spectra were reduced using a custom pipeline running standard \textsc{pyraf} procedures.

We obtained two final epochs of spectroscopic observations on 2020 July 10 and 2020 July 21, with the Gemini Multi-Object Spectrograph (GMOS-N) instrument on Gemini North. The Gemini spectra were obtained using the R400 grating ($R \approx 1900$) and a $1 \arcsec$ slit, and reduced with various tasks in the Gemini \textsc{iraf} package. 

Finally, in order to tie the spectra to an absolute flux scale, we computed synthetic photometry for the spectra using the Synthetic Magnitudes from Spectra code \citep[\textsc{sms};][]{2018MNRAS.475.1046I}. The broadband $ugrizy$ photometry described above was used to compute monochromatic fluxes that were interpolated at epochs corresponding to the spectroscopic observations, and a scaling factor was applied to the spectra where necessary.

\section{Light Curves and Explosion Parameters}

The median redshift independent distance on the NASA Extragalactic Database (NED) for the host galaxy NGC 5012 from 18 different measurements, scaled to $H_0 = 70$ km s$^{-1}$ Mpc$^{-1}$, is 41 Mpc. The three most recent Tully-Fisher (TF) distances from Spitzer 3.8$\mu$m \citep{2014MNRAS.444..527S}
and $I$-band data \citep{2016AJ....152...50T}
are between $42-43$ Mpc (for an adopted $H_0 = 75$ km s$^{-1}$), consistent with the median value stated above. The Hubble Flow distance, from the dynamically corrected recessional velocity (3169\,\kms) is also similar at $45\pm3$\,Mpc. 
We adopt the mean of the three recent TF measurements scaled to $H_0 = 70$ km s$^{-1}$ Mpc$^{-1}$, i.e. $d = 43 \pm 3$ Mpc or a distance modulus $\mu = 33.17 \pm 0.15$ for \kyg. A standard reddening law with $R_V = 3.1$ and $A_V = 0.038$ mag \citep{2011ApJ...737..103S} was adopted to correct for Galactic extinction along the line of sight. The reddening due to the host galaxy was assumed to be negligible, since \kyg\ is significantly offset from the host, and there is no sign of narrow Na~{\sc i} lines in the spectra. Thus, a total reddening of $A_V = 0.038$ mag was adopted for \kyg\ in the subsequent analysis.


\subsection{Light Curve Properties}\label{subsec:lcprop}

\begin{figure*}
	\includegraphics[width=0.8\linewidth]{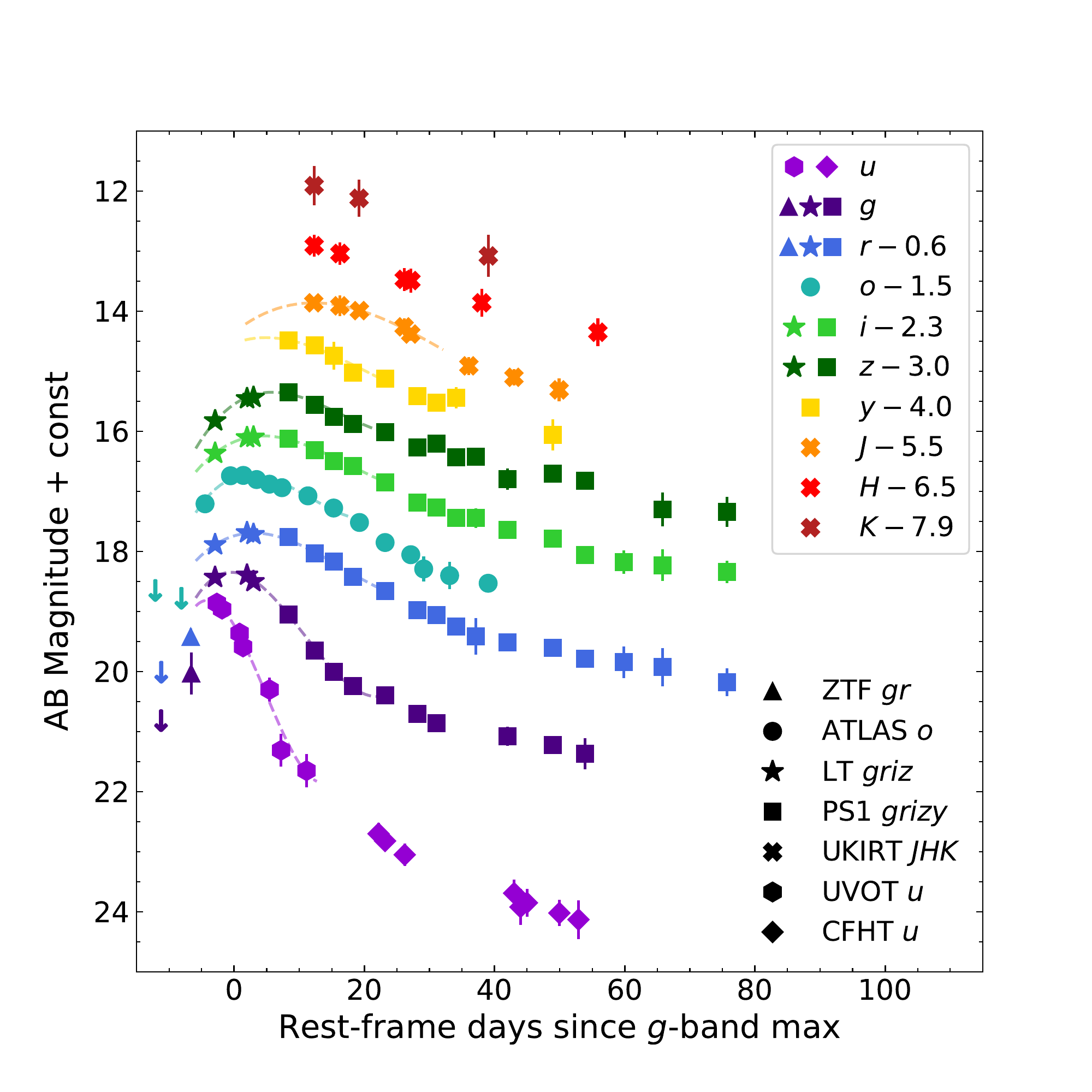}
    \caption{Multi-colour $ugrizyJHK$ light curves of SN2020kyg. All magnitudes are in the AB system. The light curves were shifted in the y-axis for clarity. ATLAS $o$-band, ZTF $g$ and $r$-band non-detections prior to discovery are also shown. The different symbols represent different observing facilities, as shown in the legend. 
    }
    \label{fig:lc}
\end{figure*}

The multi-band light curves of \kyg\ are shown in Figure~\ref{fig:lc}. The explosion epoch is well constrained, with a combination of ATLAS and Zwicky Transient Facility (ZTF) data. There are detections by ZTF \citep{2019PASP..131a8002B} two days before the ATLAS discovery, although with one being marginal it was not registered as a discovery on MJD 58991.3. The ZTF photometry 
\citep[through the Lasair broker;][]{2019RNAAS...3a..26S} reports a 3$\sigma$ detection at $g=20.03\pm0.36$ and a more secure detection at $r=20.03\pm0.14$, with ATLAS non-detections on MJDs 58985.38 ($o > 20.15$) 58987.39 ($c > 20.52$) 
and MJD 58989.35 ($o > 20.28$). 
\kyg\ was discovered a few days before peak and the pre-maximum photometric coverage is not adequate to allow a precise estimate for the explosion epoch using an expanding fireball model. The epoch of explosion was instead estimated from the best-fitting rise time from modelling the bolometric light curve (Section~\ref{subsec:bol}).




The light curve parameters, including peak absolute magnitudes, decline rates and rise times, estimated using polynomial fits, are summarised in Table~\ref{tab:lc_params}. The uncertainties on the peak luminosity were estimated by combining in quadrature the error on the peak observed magnitude from the polynomial fit, and error on distance modulus, $\mu = 33.17 \pm 0.15$ mag.
The $r$-band light curve peaks at an absolute magnitude $M_r = -14.91 \pm 0.15$,
placing it among the least luminous thermonuclear SNe observed. \kyg\ shows a modest decline rate in the $r$-band, $\Delta m_{15}(r) = 0.67 \pm 0.05$, in contrast to faint Iax events such as SN 2019gsc with $\Delta m_{15}(r) = 0.91 \pm 0.09$ \citep{2020ApJ...892L..24S}, SN 2010ae with $\Delta m_{15}(r) = 1.01 \pm 0.03$ and SN 2008ha with $\Delta m_{15}(r) = 1.11 \pm 0.04$ \citep{2014A&A...561A.146S}.
The $r$-band decline rate of \kyg\ is comparable to more luminous members of the Iax family such as SNe 2005hk \citep{2015A&A...573A...2S} and 2015H \citep{2016A&A...589A..89M} and the normal SN Ia population rather than other members of the faint Iax sample. This is in keeping with the heterogeneity seen in SN Iax light curves \citep{2016A&A...589A..89M}.

\begin{figure*}
	\includegraphics[width=0.9\linewidth]{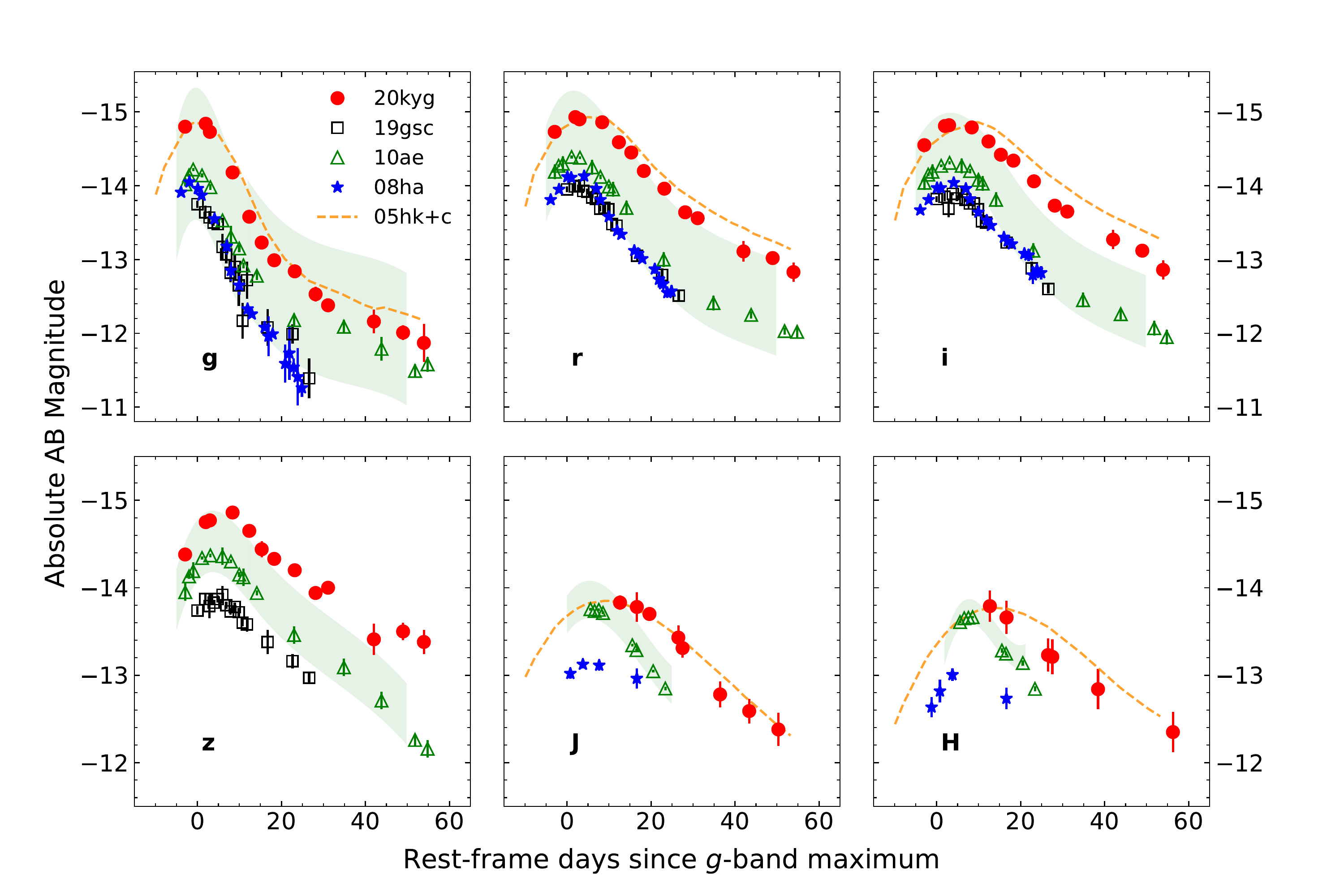}
    \caption{Absolute magnitude light curves of \kyg\ in $grizJH$ bands, compared with the Iax events SNe 2019gsc, 2010ae, 2008ha and 2005hk. For SN 2010ae, we adopt E$(B-V)_{\rm tot}=0.3$ \citep{2020ApJ...892L..24S}. The shaded region indicates the range of absolute magnitudes for SN 2010ae given the large uncertainty in the reddening, E$(B-V)_{\rm tot}=0.62 \pm 0.42$ \citep{2014A&A...561A.146S}. The light curves of SN 2005hk were shifted by $\sim 3$ magnitudes in each band to match the peak absolute magnitudes of \kyg.}
    \label{fig:grizJH}
\end{figure*}

\begin{figure*}
	\includegraphics[width=0.8\linewidth]{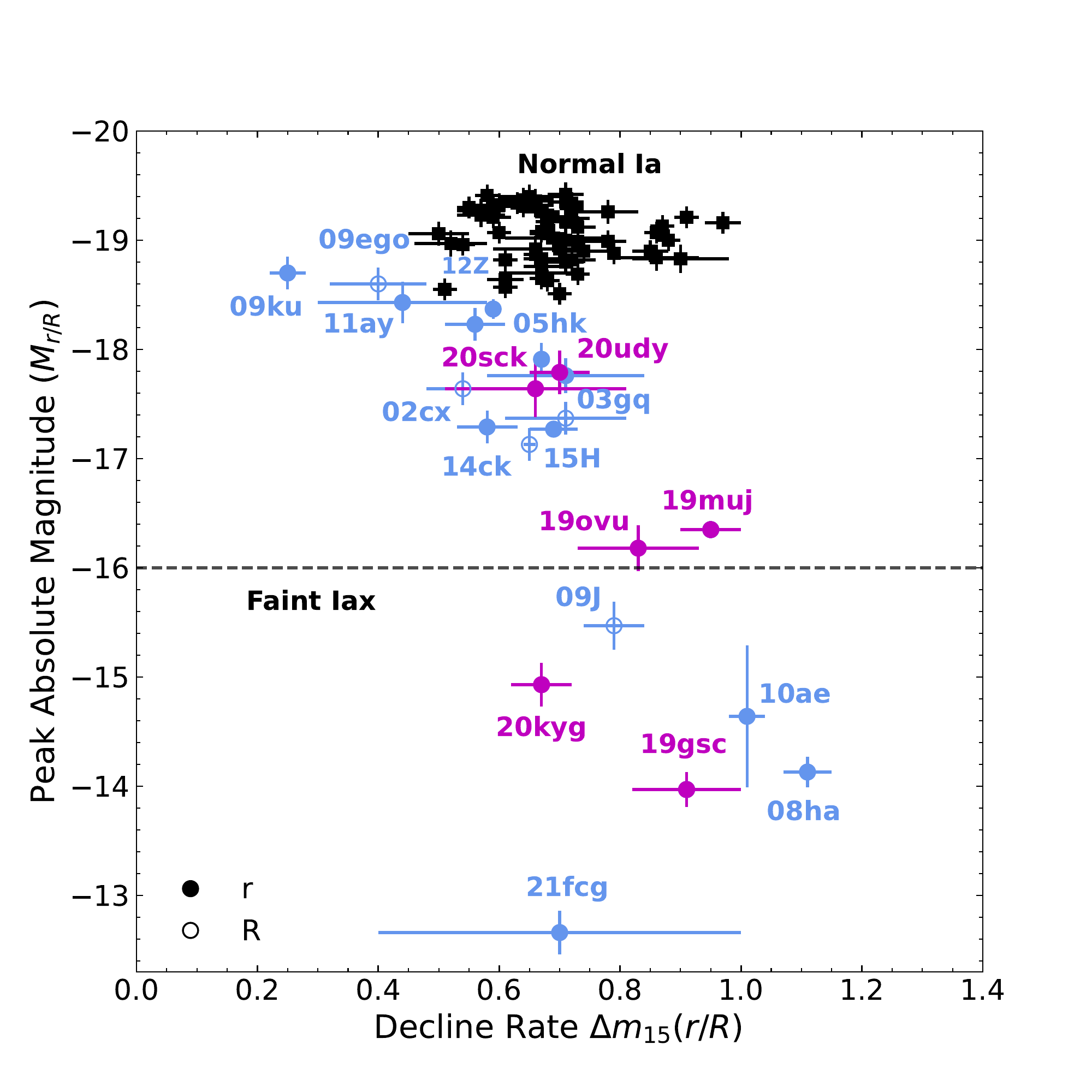}
    \caption{Peak absolute magnitude ($r$ or $R$-band) versus the decline rate parameter $\Delta m_{15}$ for well-studied SNe Iax (blue or magenta). Also shown (black) is the population of normal SNe Ia from the low redshift CSP sample \citep{2011AJ....142..156S,2017AJ....154..211K}. The SNe Iax highlighted in magenta are those from our ATLAS 100Mpc Local Volume Survey (see Section 5). 
    }
    \label{fig:PeakMag_dm15}
\end{figure*}

The absolute magnitude light curves of \kyg\ in $grizJH$ bands are shown in Figure~\ref{fig:grizJH}, along with those of Iax SNe 2019gsc \citep{2020ApJ...892L..24S}, 2010ae \citep{2014A&A...561A.146S}, 2008ha \citep{2014A&A...561A.146S} and 2005hk \citep{2007PASP..119..360P}.
The peak luminosity is known to be correlated with the decline rate in SNe Iax \citep[eg.][]{2013ApJ...767...57F,2016A&A...589A..89M,2017hsn..book..375J}, akin to the general Ia population, although the relation is much steeper in SNe Iax given the wide range in luminosity. Following \citet{2016A&A...589A..89M}, Figure~\ref{fig:PeakMag_dm15} shows the peak absolute $r$-band magnitude as a function of $r$-band decline rate for several SNe Iax, along with the normal Ia population for context. The parameters for normal SNe Ia were computed from the Carnegie Supernova Project (CSP) data \citep{2011AJ....142..156S,2017AJ....154..211K}. The $r$-band light curve for each object in the CSP sample, with adequate coverage around maximum, was fit with a polynomial function to estimate the epoch of maximum, peak observed magnitude and decline rate. The distances were derived from the redshift (scaled to $H_0 = 70$ km s$^{-1}$ Mpc$^{-1}$) and peak absolute magnitudes were computed after correcting for Galactic extinction. Photometric outliers in the CSP Ia sample were rejected by imposing the criterion $ -19.5 \leq M^{\rm peak}_r \leq -18.5$. The objects highlighted in magenta are Iax events from our ATLAS 100 Mpc Local Volume Survey (described further in Section~\ref{sec: rates}), used to estimate true volumetric rates. For the Iax events SNe 2019ovu, 2020sck and 2020udy, the peak $r$-band absolute magnitude was estimated from the ATLAS cyan and orange-band peak magnitudes, as $M_r \approx 0.35M_c + 0.65M_o$ \citep{2018PASP..130f4505T}.  


\begin{table*}
\caption{Light curve parameters for \kyg. The peak absolute magnitudes were computed for a distance modulus $\mu = 33.17 \pm 0.15$ mag and corrected for Galactic extinction along the line of sight. The uncertainties on the peak luminosity include the uncertainty on the distance modulus and the peak magnitude derived from the polynomial fit. All magnitudes are in the AB system. The rise time in each band was estimated assuming the epoch of explosion as MJD $58988.0 \pm 1.5$ (Section~\ref{subsec:bol}). 
}
\begin{tabular}{cccccc}\hline
Filter & T$_{\rm peak}$ & Peak Obs. Mag & Peak Abs. Mag & $\Delta m_{15}$ & Rise Time \\
\hline
$u$ & 58993.7 & $18.81\pm0.17$ & $-14.41\pm0.23$ & $2.81\pm0.29$ & $5.7$\\
$g$ & 58997.5 & $18.35\pm0.03$ & $-14.87\pm0.15$ & $1.53\pm0.12$ & $9.5$\\
$r$ & 59001.0 & $18.29\pm0.03$ & $-14.91\pm0.15$ & $0.67\pm0.05$ & $13.0$\\
$o$ & 59000.5 & $18.23\pm0.11$ & $-14.97\pm0.19$ & $0.68\pm0.13$ & $12.5$\\
$i$ & 59002.3 & $18.38\pm0.06$ & $-14.82\pm0.16$ & $0.57\pm0.07$ & $14.3$\\
$z$ & 59003.8 & $18.35\pm0.06$ & $-14.84\pm0.16$ & $0.58\pm0.08$ & $15.8$\\
$y$ & 59002.7 & $18.44\pm0.12$ & $-14.74\pm0.19$ & $0.53\pm0.14$ & $14.7$\\
$J$ & 59009.9 & $19.36\pm0.18$ & $-13.81\pm0.23$ & $0.49\pm0.16$ & $21.9$\\
$H$ & $-$     & $-$ & $\leq -13.77$ & $-$ & $-$\\
$K$ & $-$     & $-$ & $\leq -13.31$ & $-$ & $-$\\
\hline
\end{tabular}
\label{tab:lc_params}
\end{table*}

\subsection{Bolometric Light Curve}\label{subsec:bol}

\begin{figure*}
	\includegraphics[width=0.7\textwidth]{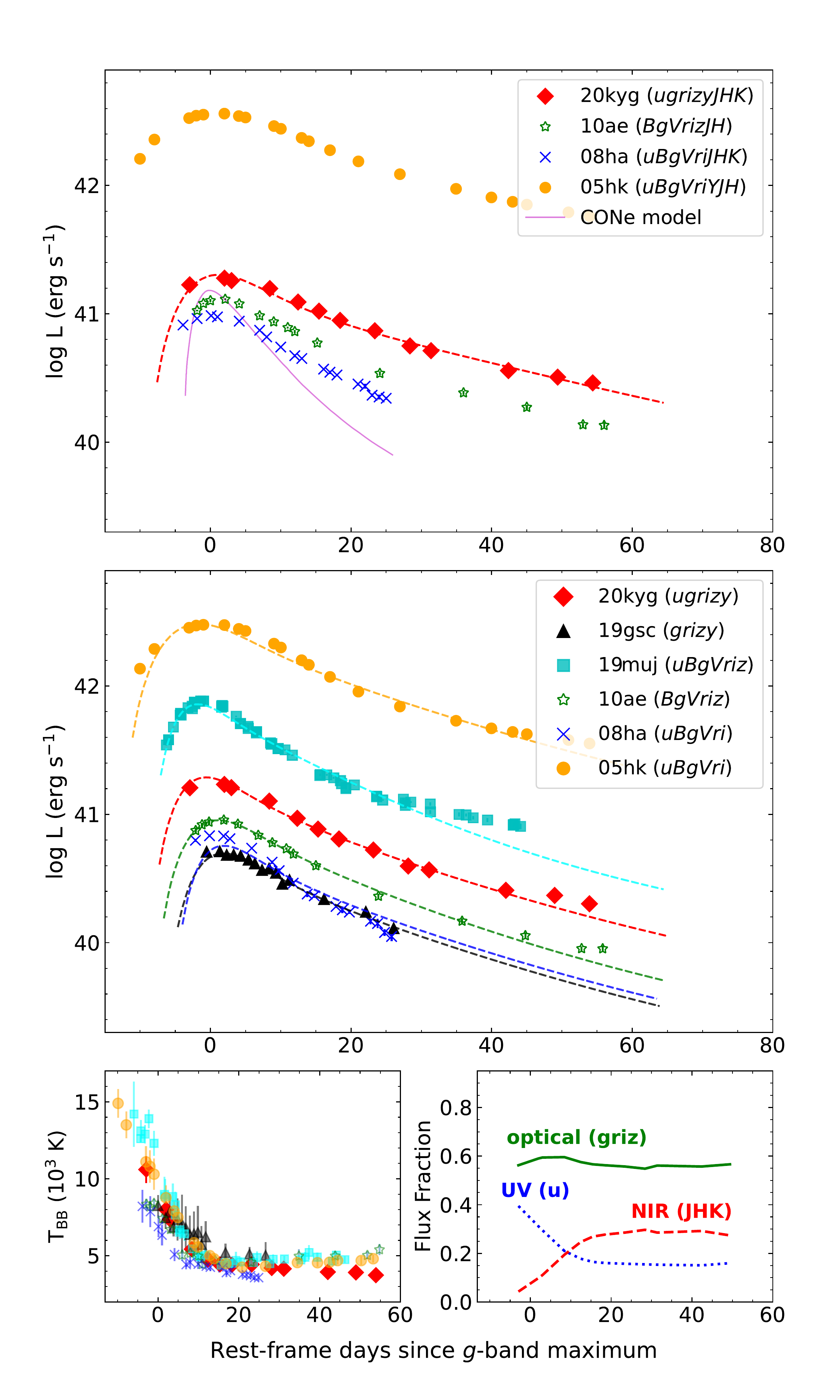}
    \caption{Top panel: quasi-bolometric light curve of \kyg\ integrated within the full observed $ugrizyJHK$ wavelength interval. The dashed line indicates the best-fitting Arnett model used to derive the explosion parameters. Shown for comparison are the quasi-bolometric light curves of Iax SNe 2010ae, 2008ha and 2005hk, all having NIR photometric coverage. The solid magenta line is the angle averaged synthetic bolometric light curve for the hybrid CONe WD deflagration model \citep{2015MNRAS.450.3045K}. Middle Panel: UV-optical quasi-bolometric light curve of \kyg\ integrated within $ugrizy$ bands, along with quasi-bolometric light curves of Iax SNe 2019gsc \citep{2020ApJ...892L..24S}, 2019muj \citep{2021MNRAS.501.1078B}, 2010ae \citep{2014A&A...561A.146S}, 2008ha \citep{2009AJ....138..376F,2014A&A...561A.146S} and 2005hk \citep{2007PASP..119..360P}, integrated within similar wavelength intervals for a direct comparison. Also shown are the best-fitting Arnett models for each SN. Bottom left panel: evolution of the derived blackbody temperature for \kyg\ and the comparison sample from fitting the SED. Bottom right panel: evolution of fractional contribution of the UV ($u$), optical ($griz$) and NIR ($JHK$) bands to the bolometric flux for \kyg.}
    \label{fig:bolometric}
\end{figure*}

The multi-colour $ugrizyJHK$ photometry of \kyg\ was used to compute the integrated quasi-bolometric flux using the \texttt{SuperBol} code \citep{2018RNAAS...2d.230N}. In addition to the quasi-bolometric flux integrated within the limits defined by the photometric passbands, \texttt{SuperBol} also computes a full blackbody integration from a fit to the spectral energy distribution (SED) to account for missing flux. However, due to heavy line blanketing in the UV regime, a full blackbody extrapolation would likely overestimate the bolometric flux. 
The full $ugrizyJHK$ (top panel) and the $ugrizy$ (middle panel) quasi-bolometric light curve of \kyg\ is shown in Figure~\ref{fig:bolometric}, along with the quasi-bolometric light curves of the faint Iax SNe 2008ha \citep{2009AJ....138..376F,2014A&A...561A.146S}, 2010ae \citep{2014A&A...561A.146S}, 2019gsc \citep{2020ApJ...892L..24S}, the intermediate luminosity Iax SN 2019muj \citep{2021MNRAS.501.1078B}, and the luminous Iax SN 2005hk \citep{2007PASP..119..360P} for comparison. 
The quasi-bolometric light curves of all the objects  were computed using the same method for consistency and direct comparison. 
The magnitudes for each SN were corrected for extinction assuming a standard reddening law with $R_V=3.1$ prior to computing the bolometric fluxes. The total $E(B-V)$ values adopted were 0.09 for SN 2005hk \citep{2007PASP..119..360P}, 0.08 for SN 2008ha \citep{2009AJ....138..376F}, 0.30 for SN 2010ae \citep{2020ApJ...892L..24S}, 0.02 for SN 2019muj \citep{2021MNRAS.501.1078B} and 0.01 for SN 2019gsc \citep{2020ApJ...892L..24S}. Also shown in Figure~\ref{fig:bolometric} is the evolution of effective blackbody temperature computed from the fit of a Planck function to the 
observed object SEDs (photometry available in all passbands was used for the fit). 
The blackbody temperature evolution of all these objects is quite similar, despite the large range in peak luminosity, 
indicating that it is the radius of the emitting photosphere, and therefore the expansion velocity that is the main driver of the luminosity diversity. 
The contribution of NIR $JHK$ bands to the quasi-bolometric flux for \kyg\ is $\sim 4\%$ at $-3$d, rising steadily to $\sim 30\%$ at $+30$d. The $u$-band contribution on the other hand drops from $\sim 40\%$ at $-3$d to $\sim 15\%$ at $+30$d. The time-dependent fractional contribution of the optical ($griz$), UV ($u$) and NIR ($JHK$) bands to the bolometric flux is shown in Figure~\ref{fig:bolometric}.

In order to derive explosion parameters such as the $^{56}$Ni mass ($M_{\rm Ni}$), total ejecta mass ($M_{\rm ej}$) and kinetic energy ($E_{\rm k}$), we fit the bolometric light curves of \kyg\ and other SNe Iax with an Arnett model \citep{1982ApJ...253..785A} using the analytical treatment formulated by \citet{2008MNRAS.383.1485V}. The opacity was fixed at $\kappa_{\rm opt} = 0.07$ cm$^2$ g$^{-1}$. For a homogeneous ejecta density with a photospheric velocity of $v_{\rm ph}$, the kinetic energy of the explosion \citep{1982ApJ...253..785A} can be expressed as 
$$E_{\rm k} \approx \frac{3}{5} \frac{M_{\rm ej}{v_{\rm ph}^2}}{2}$$
We use the constraints from spectroscopic observations to fix $v_{\rm ph}$. For \kyg, $v_{\rm ph}$ was fixed at $4500$ \kms, the \SiIIa\ velocity measured from the $-3.1$d spectrum. 
The free parameters in the fitting procedure are therefore $M_{\rm Ni}$, $M_{\rm ej}$ and the rise time $t_{\rm R}$. Fitting the $ugrizyJHK$ quasi-bolometric light curve of \kyg\ with this model yields best-fit parameters of $M_{\rm Ni} = 0.007 \pm 0.001$ \msun, 
$M_{\rm ej} = 0.36^{+0.08}_{-0.06}$ \msun, and $t_{\rm R} = 9.8 \pm 1.5$ days. This $g$-band rise time implies the epoch of explosion was MJD $58988.0 \pm 1.5$, consistent with the pre-discovery constraints from ATLAS and ZTF (Section~\ref{subsec:lcprop}).
The kinetic energy of the explosion is then
$E_{\rm k}$=$4.4^{+2.2}_{-1.5}\times 10^{49}$ erg. 
The uncertainties were estimated by computing the best-fit parameters for a range of $v_{\rm ph}$ values between $4000$ and $5000$ \kms. The $M_{\rm Ni}$ is a factor of $\sim 100$ lower than that for normal SNe Ia. However, the inferred $M_{\rm ej}$ is only a factor of a few lower than \mch. This extreme ratio of $M_{\rm Ni}/M_{\rm ej}$ is a feature of faint SNe Iax that makes it challenging for  a theoretical interpretation involving a thermonuclear explosion. The disparity  is illustrated in the narrow light curve of the \citet{2015MNRAS.450.3045K} model compared to the data in Figure~\ref{fig:bolometric}.



\section{Spectral Evolution}

\begin{figure*}
\includegraphics[width=0.9\textwidth]{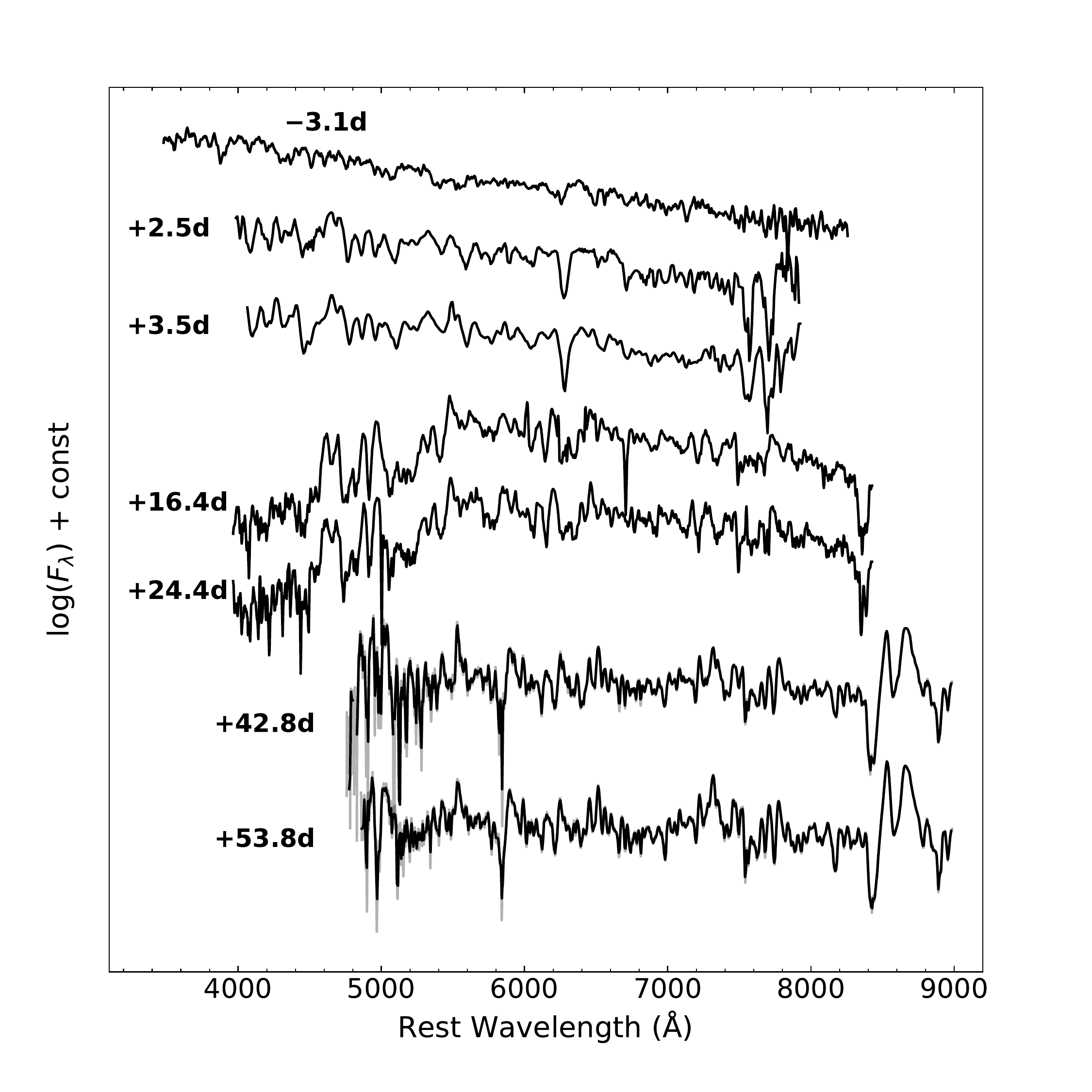}
\caption{Spectral evolution of \kyg\ between $-3$ to $+54$ days relative to the epoch of $g$-band maximum.
}
\label{fig:specevolution}
\end{figure*}

The spectral evolution of \kyg\ between $-3$ to $+54$ days relative to \gmax\ is shown in Figure~\ref{fig:specevolution}. A summary of spectroscopic observations is presented in Table~\ref{tab:spec_log}. Apart from the lower velocities, spectra of SNe Iax around maximum are generally similar to normal SNe Ia and bluer 1991T-like SNe, showing rather weak Si~{\sc ii} and prominent Fe~{\sc iii} \citep{2013ApJ...767...57F,2017hsn..book..375J}. SN Iax spectra start diverging from normal SN Ia spectra at later epochs, showing a combination of forbidden emission lines and permitted P-Cygni lines persisting for over a year after explosion \citep{2006AJ....132..189J,2014ApJ...792...29F}.
The low observed velocities in SNe Iax, especially in faint Iax, lead to spectra with narrow, resolved lines that are otherwise blended in normal SNe Ia \citep{2013ApJ...767...57F}, resulting in complex spectra rich in features.

\begin{table}
    \centering
    \caption{Log of spectroscopic observations for SN 2019gsc. The phase is relative to the epoch of \gmax\ in the SN rest frame.}
    \resizebox{\columnwidth}{!}{
    \begin{tabular}{ccccc}
    \hline
    Date         & MJD      & Phase   & Instrument    & Exposure        \\
    (yyyy/mm/dd) &          & (days)  &               & (s)             \\ \hline
    2020/05/25   & 58994.4 & $-3.1$  & FTN/FLOYDS-N  & 3600 \\
    2020/05/30   & 59000.0 & +2.5    & LT/SPRAT      & 2100 \\
    2020/05/31   & 59001.0 & +3.5    & LT/SPRAT      & 2100  \\
    2020/06/13   & 59013.9 & +16.4   & NOT/ALFOSC    & $3 \times 900$ \\
    2020/06/21   & 59021.9 & +24.4   & NOT/ALFOSC    & $3 \times 900$ \\
    2020/07/10   & 59040.3 & +42.8  & Gemini/GMOS-N & $4 \times 350$ \\
    2020/07/21   & 59051.3 & +53.8  & Gemini/GMOS-N & $4 \times 580$ \\
    \hline
    \end{tabular}
    }
    \label{tab:spec_log}
\end{table}

\begin{figure*}
\includegraphics[width=\linewidth]{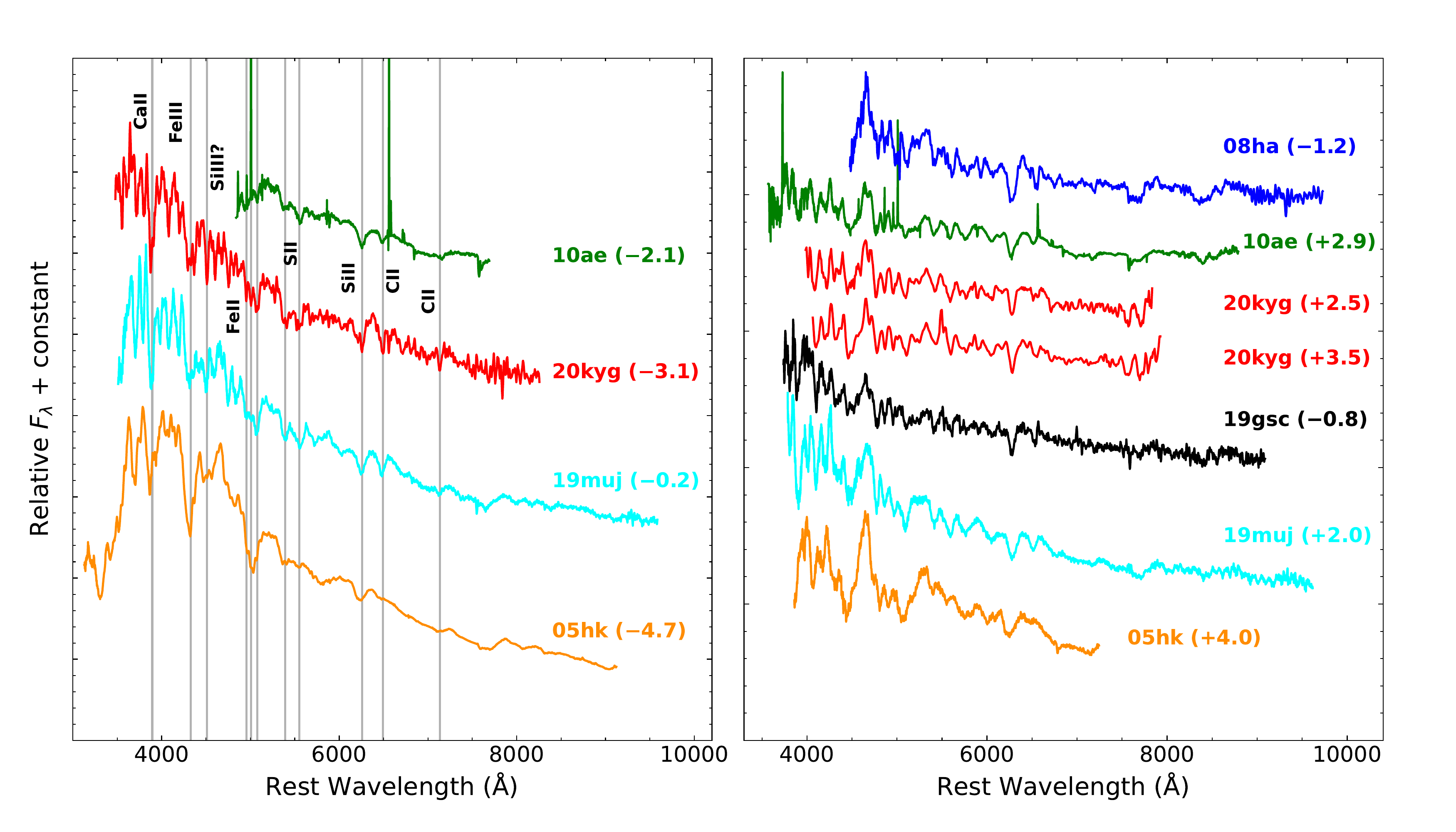}
\caption{Spectral comparison of \kyg\ with spectra of faint Iax events SNe 2008ha \citep{2009Natur.459..674V}, 2010ae \citep{2014A&A...561A.146S} and 2019gsc \citep{2020ApJ...892L..24S}, the intermediate luminosity Iax SN 2019muj \citep{2021MNRAS.501.1078B} and the luminous Iax SN 2005hk  \citep{2006PASP..118..722C,2007PASP..119..360P}. Phases are in days relative to $g$-band maximum for SNe 2020kyg and 2019gsc, and relative to $B$-band maximum for the rest.}
\label{fig:spec_comp1}
\end{figure*}

The early spectra of \kyg\ are characterised by a blue continuum and narrow P-Cygni features of intermediate mass elements (IMEs) and
iron group elements (IGEs). The spectra are compared to the faint Iax SNe 2008ha \citep{2009Natur.459..674V}, 2010ae \citep{2014A&A...561A.146S} and 2019gsc \citep{2020ApJ...892L..24S}, the intermediate luminosity Iax SN 2019muj \citep{2021MNRAS.501.1078B} and the luminous Iax SN 2005hk \citep{2006PASP..118..722C,2007PASP..119..360P} in Figure~\ref{fig:spec_comp1}. The spectra at these epochs are overall quite similar, although in SN 2005hk the Si~{\sc ii} features are shallower and the \CIIa\ feature is less prominent. The \CIIb\ feature is also clearly detected in the $-3.1$d spectrum of \kyg. In the pre-maximum phase (Figure~\ref{fig:spec_comp1}, left panel), C~{\sc ii} is particularly conspicuous (except for SN 2005hk), being comparable in strength to Si~{\sc ii} $\lambda 6355$. Additionally, the `W' feature around 5500\AA, attributed to blended features of S~{\sc ii} $\lambda 5454, 5640 \AA$ \citep{2006MNRAS.370..299H}, tends to be stronger in fainter SNe Iax. \citet{2020ApJ...892L..24S} couldn't adequately reproduce this feature in their \tardis\ models for SN 2019gsc with a reasonable sulphur abundance of $25-50 \%$ of the silicon mass fraction \citep[eg.][]{2015MNRAS.450.3045K}. It is possible that other species like Cr~{\sc ii} and Sc~{\sc ii} are  contributing to this feature in the lower luminosity Iax events. 

The $-3.1$d spectrum also shows an absorption trough at $\sim 4500$\AA\, that we tentatively identify as a blend of Si~{\sc iii} $4553, 4568, 4575$\AA\ lines.
This doubly ionised Si feature, indicative of a hot photosphere, has been identified in pre-maximum spectra of SN 1991T \citep{2014MNRAS.445..711S}. Si~{\sc iii} features have also been identified previously in SNe Iax, such as in the optical spectra of SN 2005hk \citep{2008ApJ...680..580S} and the NIR spectra of SN 2014ck \citep{2016MNRAS.459.1018T}.
The C~{\sc ii} features, although very prominent in the $-3.1$d spectrum, are weak or absent in the +2.5 and +3.5d spectra of \kyg\ (Figure~\ref{fig:spec_comp1}, right panel). The features around 6500\AA\ in the +2.5 and +3.5d spectra thus likely have contribution from Co~{\sc ii} \citep{2015MNRAS.453.2103S,2015MNRAS.449.3581J,2016MNRAS.459.1018T}.
In general, SNe Iax tend to display stronger C~{\sc ii} features in pre-maximum spectra relative to normal SNe Ia at similar phases as shown by \citet{2013ApJ...767...57F}, who found signatures of C~{\sc ii} in $>82\%$ of their Iax sample. 

\begin{figure*}
\includegraphics[width=\textwidth]{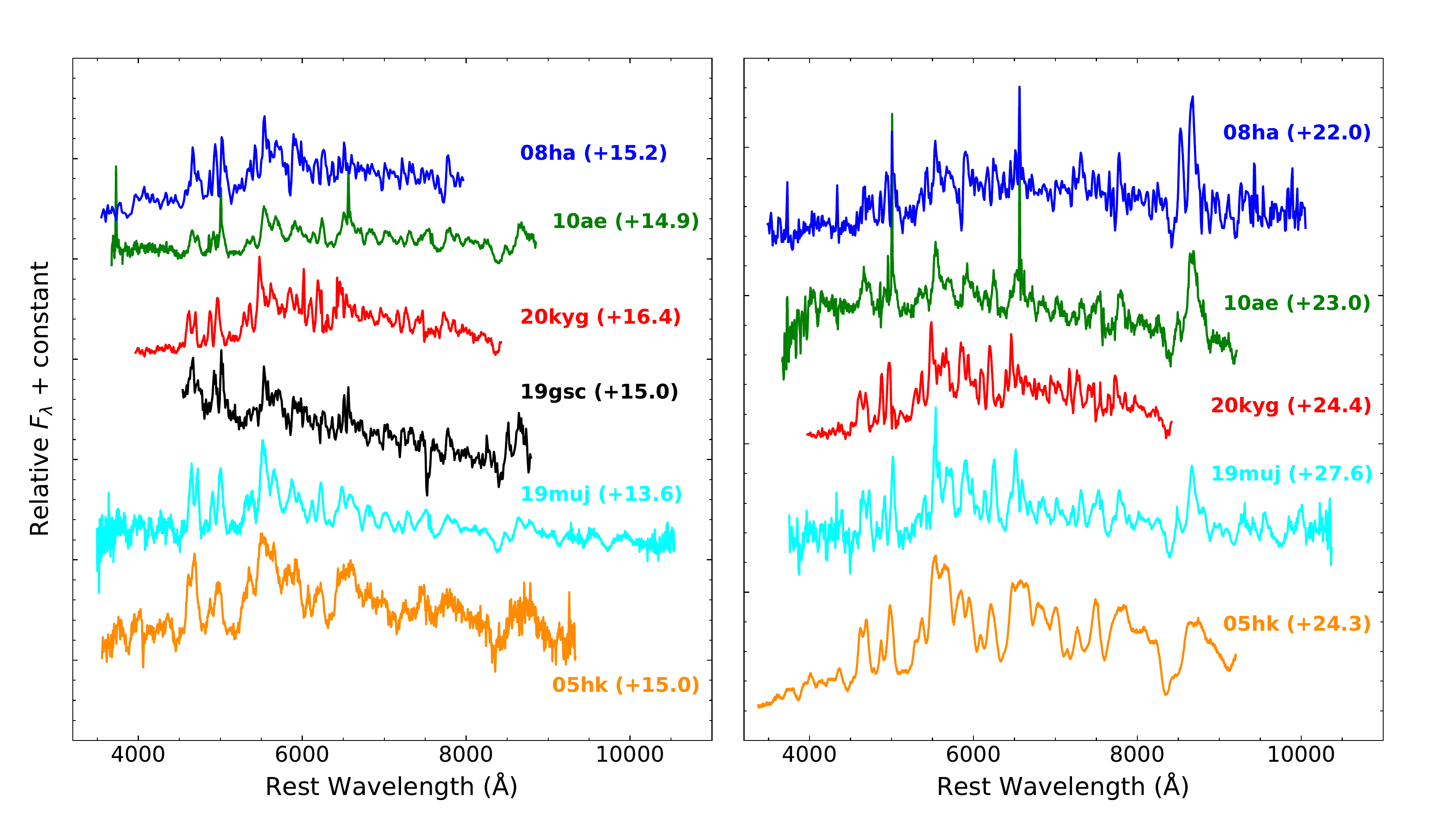}
\caption{Spectral comparison of \kyg\ with SNe 2008ha \citep{2009Natur.459..674V,2009AJ....138..376F}, 2010ae \citep{2014A&A...561A.146S}, 2019gsc \citep{2020ApJ...892L..24S}, 2019muj \citep{2021MNRAS.501.1078B} and 2005hk \citep{2007PASP..119..360P}. Phases are in days relative to $g$-band maximum for SNe 2020kyg and 2019gsc, and relative to $B$-band maximum for the others. 
}
\label{fig:spec_comp2}
\end{figure*}

\begin{figure*}
\includegraphics[width=0.8\textwidth]{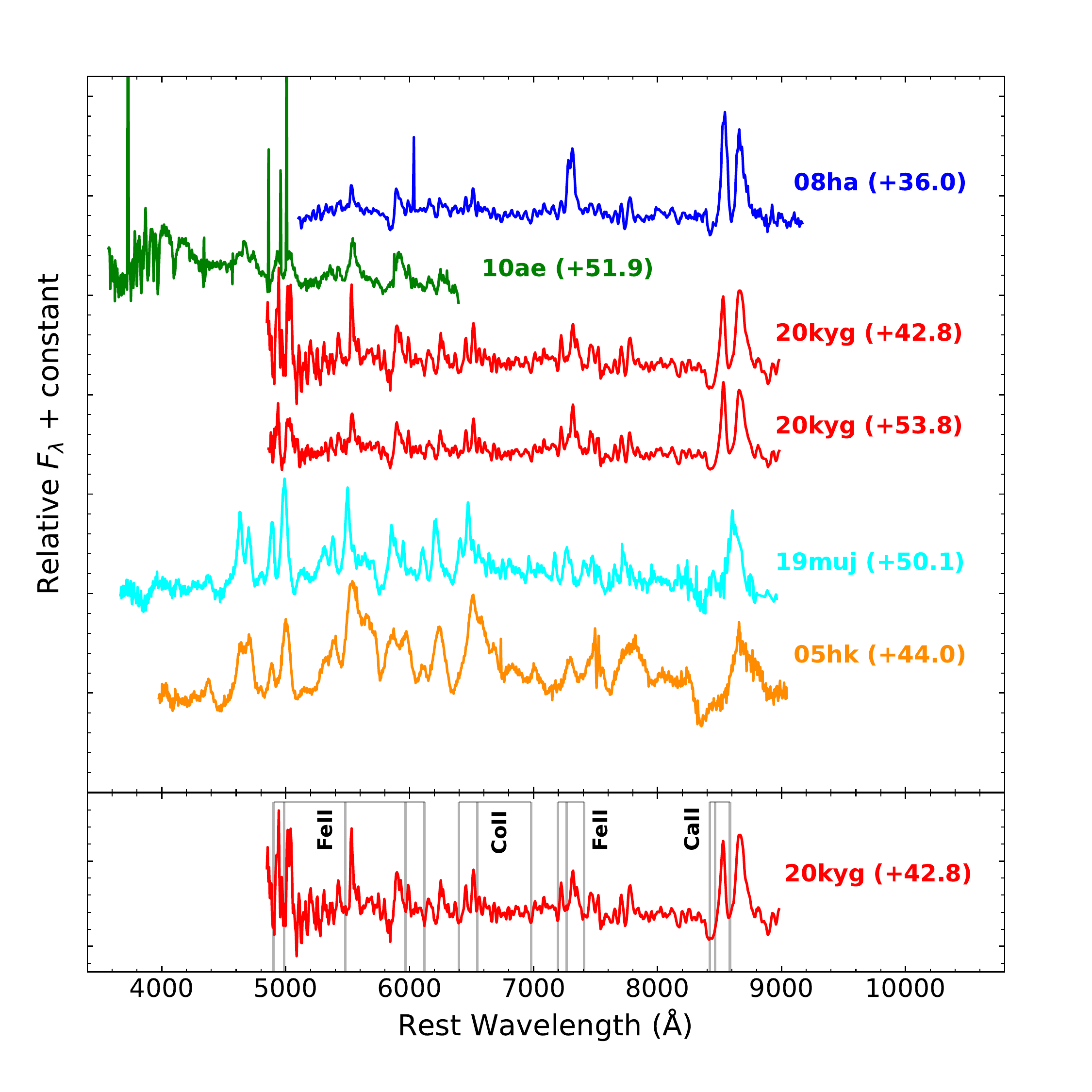}
\caption{+42.8 and +53.8d spectra of \kyg\ with spectra of SNe 2008ha \citep{2009Natur.459..674V}, 2010ae \citep{2014A&A...561A.146S}, 2019muj \citep{2021MNRAS.501.1078B} and 2005hk  \citep{2007PASP..119..360P} at similar epochs for comparison. Phases are in days relative to $g$-band maximum for SNe 2020kyg and 2019gsc, and relative to $B$-band maximum for the others.}
\label{fig:spec_comp3}
\end{figure*}

Figures~\ref{fig:spec_comp2} and \ref{fig:spec_comp3} show the spectra of \kyg\ between +16 to +54d, along with spectra of the comparison sample. At these later epochs, the spectra are increasingly dominated by permitted and forbidden features of Fe~{\sc ii}, Co~{\sc ii} and Ca~{\sc ii}. The \SiIIa\ feature is weak in the +16.4 and +24.4d spectra. The Ca~{\sc ii} NIR triplet is prominent in the +42.8 and +53.8d spectra, similar to SN 2008ha. These later epochs also show the emergence of forbidden [Ca~{\sc ii}] $\lambda \lambda 7292,7324$ emission, although this feature is not as pronounced as in the case of SN 2008ha, where this feature is already quite strong in the +36d spectrum. The +54d spectrum of SN 2019gsc obtained by \citet{2020MNRAS.496.1132T} showed a strong forbidden Ca~{\sc ii} $\lambda \lambda 7292,7324$ doublet, very similar to SN 2008ha. This highlights a more rapid evolution in SNe 2008ha and 2019gsc . 

\subsection{Expansion Velocity}

\begin{figure}
\includegraphics[width=\linewidth]{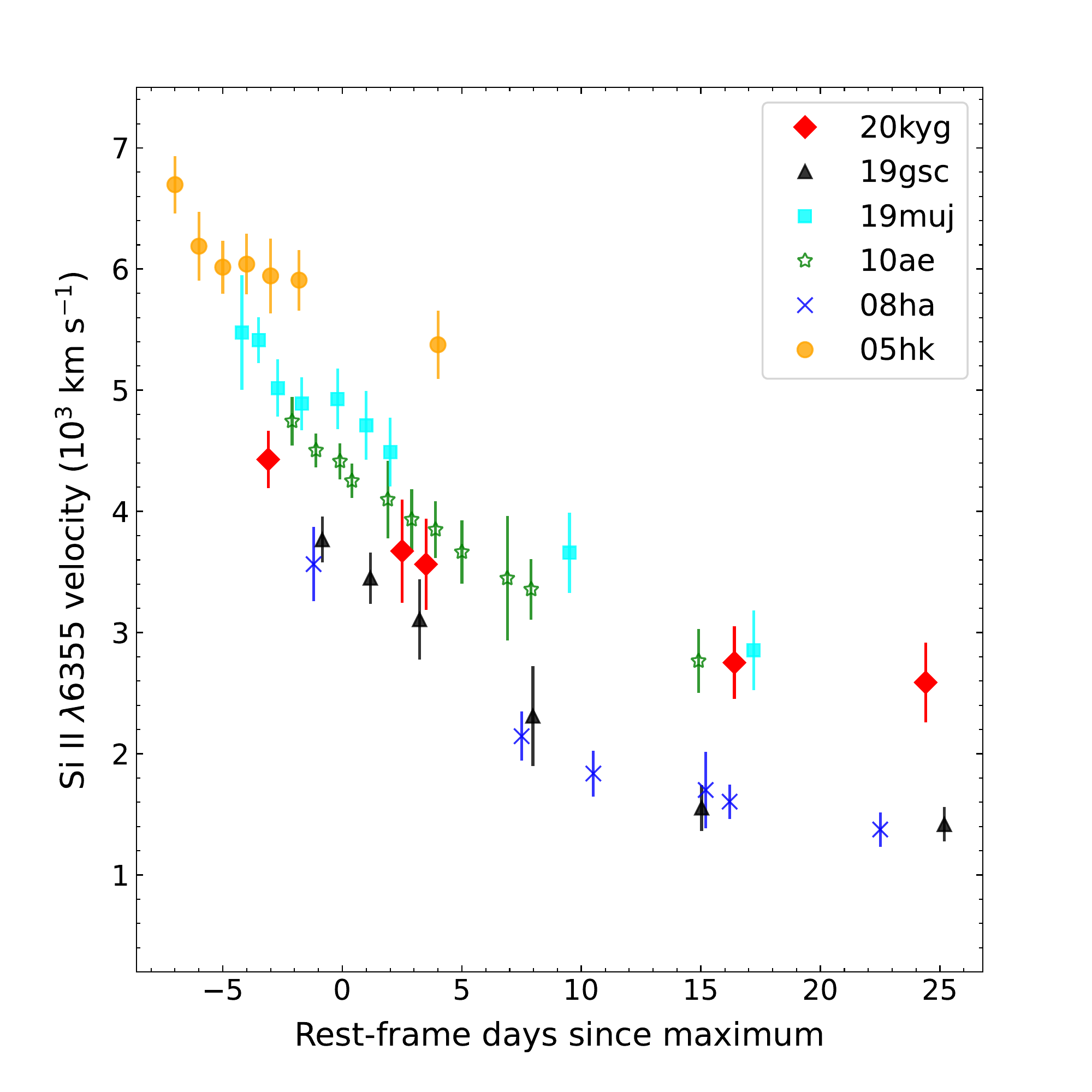}
\caption{Evolution of the measured \SiIIa\ velocity for \kyg, compared with that of SNe 2019gsc, 2019muj, 2010ae, 2008ha and 2005hk.}
\label{fig:SiIIvel}
\end{figure}

The \SiIIa\ velocity evolution of \kyg\ and other Iax in our comparison sample is shown in Figure~\ref{fig:SiIIvel}.
The \SiIIa\ velocity of \kyg\ evolves from $\sim 4400$ \kms\ at $-3.1$d to $\sim 3600$ \kms\ at $+3.5$d. The \CIIa\ velocity at $-3.1$d is $\sim 4000$ \kms, lower than the Si velocity. The expansion velocities were deduced using a Gaussian fit routine implemented using \texttt{specutils}. The ratio between the \CIIa\ and \SiIIa\ velocities at pre-maximum epochs maximum is known to be slightly above unity in normal SNe Ia \citep{2011ApJ...732...30P,2012ApJ...745...74F}, with both features showing a parallel evolution and the C~{\sc ii} velocity typically staying $\sim 1000$ \kms\ higher. This is consistent with a spherically symmetric explosion where unburnt carbon would be expected in the outer layers of the ejecta. This velocity ratio is $\sim 0.9$ for \kyg, and has been observed to be lower than unity for most SNe Iax that have pre-maximum spectra with reliable C~{\sc ii} identifications \citep[eg.][]{2020MNRAS.496.1132T}. Interestingly, this ratio is much lower ($0.5-0.6$) in the low luminosity Iax events SNe 2008ha and 2019gsc \citep{2020MNRAS.496.1132T}. \citet{2011ApJ...732...30P} interpreted this as an effect of asymmetry in the ejecta, with carbon distributed in clumps not along the line of sight.

Although a correlation between peak luminosity and ejecta velocity, with more luminous SNe Iax showing higher ejecta velocities has been suggested \citep{2010ApJ...720..704M,2013ApJ...767...57F}, notable outliers like  SN 2009ku \citep{2011ApJ...731L..11N} and SN 2014ck \citep{2016MNRAS.459.1018T} would argue against a single parameter description of this family. 

\subsection{Spectral Modeling}

We use the fast 1D Monte Carlo radiative transfer code \tardis\ \citep{2014MNRAS.440..387K,tardis2} to model the early photospheric spectra of \kyg. \tardis\ assumes a sharp inner boundary or photosphere emitting a blackbody continuum. 
The expanding SN ejecta above this inner boundary is divided into spherical shells with user-defined density and abundance profiles. The synthetic spectrum is computed based on the interaction of virtual photon packets with the SN ejecta. \tardis\ has been effectively used in the literature to model Iax spectra and constrain the ejecta composition and velocities \citep[eg.][]{2016A&A...589A..89M,2017A&A...601A..62M,2017MNRAS.471.4865B,2018MNRAS.480.3609B,2021MNRAS.501.1078B,2020ApJ...892L..24S}. 
The aim of the empirical modeling is to arrive at a self-consistent model for the SN ejecta that reproduces the primary features of the photospheric spectra. Given the degeneracy between several parameters in the \tardis\ model, such as central density, density profile, inner and outer velocity boundaries, mass fractions of various species etc., our model is not necessarily unique. The only parameters that were varied for an individual epoch were the time-dependent parameters, i.e. the time since explosion, inner velocity boundary of the computation volume, input luminosity (derived from the observed bolometric light curve), and mass fractions of radioactive isotopes. The outer velocity boundary was fixed at 8000 \kms. The time since explosion for each spectrum was estimated assuming the explosion epoch as MJD 58988.0, derived from the best-fit rise time from modeling the bolometric light curve (Section~\ref{subsec:bol}). 
For the ejecta, we use the simplest case of a uniform abundance distribution. For SN 2019muj, \citet{2021MNRAS.501.1078B} noted that a stratified abundance did not significantly improve the fits over a uniform abundance. An exponential profile was adopted for the density, similar to SN 2019gsc \citep{2020ApJ...892L..24S}, with central density $\rho_0 = 6\times 10^{-12}$ g cm$^{-3}$, $t_0 = 2$ days, and $v_0 = 3000$ \kms, where the density is a function of time since explosion ($t$) and velocity ($v$):
$$\rho(v,\, t) = \rho_0 (\frac{t_0}{t})^3 \mathrm{exp}(-v/v_0)$$

Our model is primarily composed of carbon, oxygen and neon, with contribution from silicon, sulphur, magnesium, calcium, chromium and IGEs (Table~\ref{tab:composition}). We use neon as a `filler' element to reduce the oxygen content, since the O~{\sc i} $\lambda 7774$ feature is not prominent in the observed spectra at these epochs, and adding $\sim 20\%$ of neon does not produce any additional unwanted features in the synthetic spectra.
The observed and synthetic spectra for $-3.1$, $+2.5$ and $+3.5$d are shown in Figure~\ref{fig:tardisfit}. The \SiIIa\ feature is relatively weak in the $-3.1$d spectrum, adequately reproduced by 1\% silicon in the \tardis\ model, whereas the $+2.5$d and $+3.5$d spectra require 3\% silicon. Therefore, we use a silicon abundance of 2\% to fit the three spectra together. The `W' feature at $\sim 5500$\AA\ attributed to S~{\sc ii} is clearly underrepresented in the $-3.1$d synthetic spectrum. The predicted sulphur yield in deflagration simulations is typically a fraction of the silicon abundance. For example, the \citet{2015MNRAS.450.3045K} deflagration simulation predicts a sulphur to silicon ratio of 0.44. The sulphur abundance in our model is 1\%, whereas the silicon abundance is 2\%. Increasing the sulphur abundance to 10\% improves the fit to the red wing of the `W' feature, although the blue wing is still under-fit (Figure~\ref{fig:tardisfit}, inset). However such a high sulphur to silicon abundance would be difficult to reconcile with theoretical predictions. 
\kyg\ shows no obvious signs of helium in the spectra. We therefore do not include helium in our model, although \citet{2019A&A...622A.102M} tentatively identified helium in the NIR spectrum of the faint Iax SN 2010ae.

The next observed spectrum after $+3.5$d is at $+16.4$d ($\sim 26$ days past explosion). The +16.4d observed spectrum is already dominated by features of IGEs (Figure~\ref{fig:spec_comp2}), and the continuum shape suggests the ejecta is transitioning to an optically thin phase. The same \tardis\ model computed at this epoch produces a poor fit to the continuum and features of the data. Pure deflagrations are expected to produce thoroughly mixed ejecta, suggesting that radioactive material is likely to be present in the outer layers \citep{2013MNRAS.429.2287K}. \tardis\ assumes the material within the computation volume to be in radiative equilibrium, i.e. there are no radioactive energy sources above the inner velocity boundary. This assumption becomes progressively poor as the photosphere recedes within the $^{56}$Ni rich layers of the ejecta \citep{2014MNRAS.440..387K}, especially in the pure deflagration scenario where $^{56}$Ni is already expected to be present in the outer layers. This might explain why our model produces a poor match to the +16.4d spectrum. 

\begin{figure}
\includegraphics[width=\linewidth]{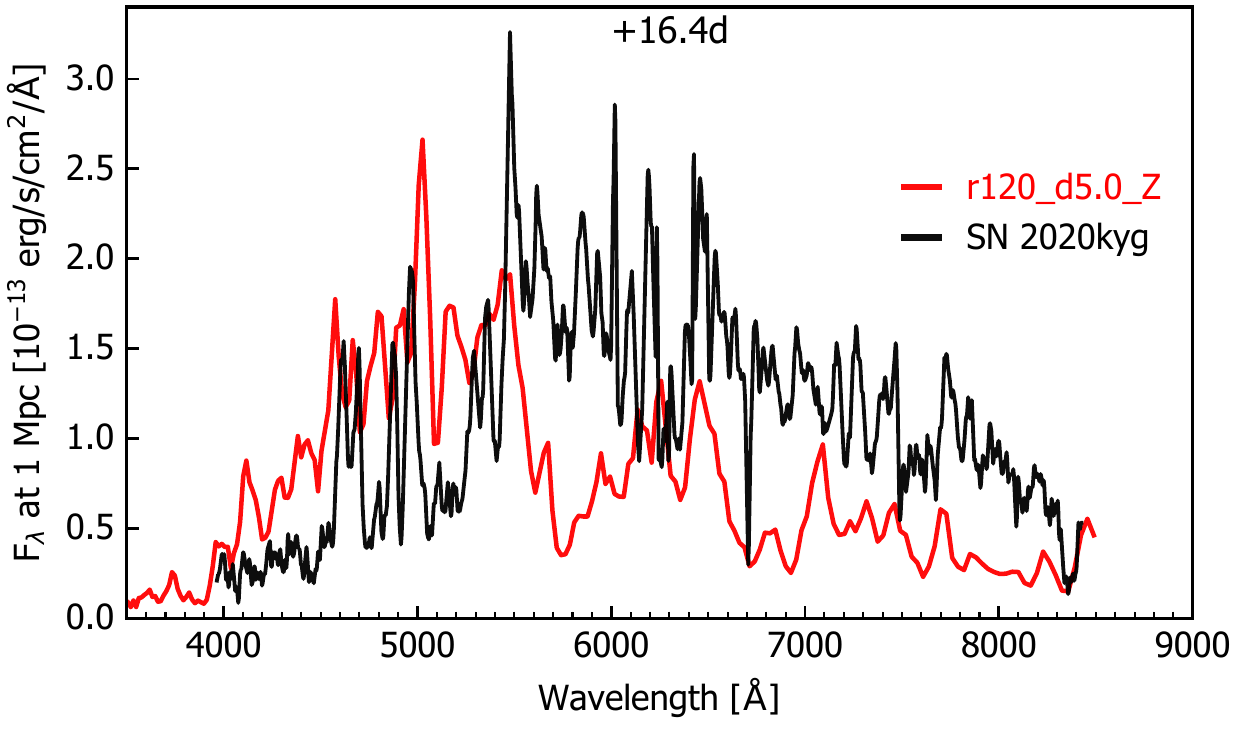}
\caption{+16.4d spectrum of \kyg\, compared to a synthetic spectrum computed using \artis, corresponding to the \texttt{r120\textunderscore d5.0\textunderscore Z} model of \citet{2021arXiv210902926L}.} 
\label{fig:artiscomp}
\end{figure}

\begin{figure*}
\includegraphics[width=0.8\linewidth]{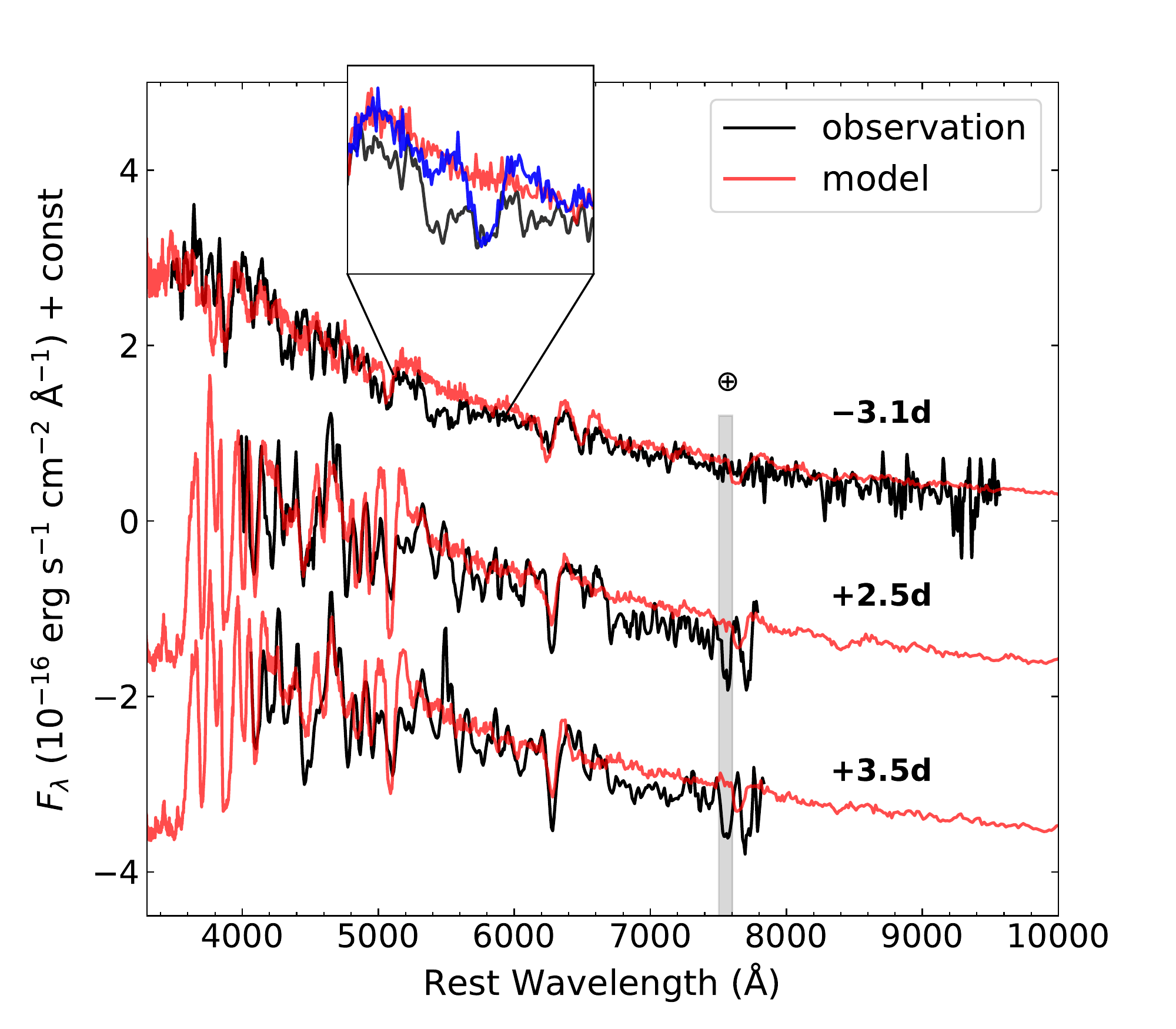}
\caption{Early photospheric spectra of \kyg\ at $-3.1$, +2.5 and +3.5d (black), shown with the \tardis\ models in red. The y-axis represents $F_{\lambda}$ in units of $10^{-16}$ erg s$^{-1}$ cm$^{-2}$ \AA$^{-1}$ for the $-3.1$d observed and synthetic spectra, and an offset was applied to the $+2.5$d and $+3.5$d spectra for clarity. Inset shows how increasing sulphur abundance to 10\% (blue line) improves the fit to the `W' feature.
}
\label{fig:tardisfit}
\end{figure*}

We compare our +16.4d spectrum to synthetic spectra generated using detailed radiative transfer calculations, following 3D deflagration simulations of \mch\ CO WDs explored by \citet{2021arXiv210902926L}. The synthetic spectra were computed using the \artis\ code \citep{2007MNRAS.375..154S,2009MNRAS.398.1809K}. The \texttt{r120\textunderscore d5.0\textunderscore Z} model \citep{2021arXiv210902926L} was chosen for the comparison. With $M_r \approx -15.6$, the model is more luminous than \kyg\ at peak, but it evolves faster and the luminosity is comparable to \kyg\ at +16d. Unlike the \tardis\ modeling, this is a forward model and no tuning or scaling was applied to the synthetic spectrum for comparing with \kyg\ (Figure~\ref{fig:artiscomp}). Although the agreement is poor, the overall continuum shape is reproduced better than the \tardis\ fit at this epoch. This is expected since \artis\ performs an approximate non-local thermodynamic equilibrium (NLTE) treatment of the plasma.

\begin{table*}
    \centering
    \caption{Mass fractions of different chemical elements and other parameters used to generate the synthetic \tardis\ spectra models for \kyg. $v_{\rm inner}$ denotes the inner boundary of the computation volume and $t$ is the time since explosion. The emergent luminosity ($L$) was fixed by interpolating the bolometric light curve at the relevant epochs.}
    \begin{tabular}{cccccccccccc}
    \hline
    $t$ & $L$ & $v_{\rm inner}$ & $X$(C) & $X$(Ne) & $X$(O) & $X$(Si) & $X$(S) & $X$(Fe) & $X$(Co) & $X$(Ni) & $X$(Ca)\\
    (days) & ($\log L/L_{\odot}$) & (km s$^{-1}$) & & & & &  & & & \\ \hline
    6.5 & 7.80 & 4200 &  0.60 & 0.20 & 0.17 & 0.02 & 0.01 & $10^{-3}$ & $10^{-4}$ & $10^{-4}$ & $4\times 10^{-5}$\\
    12.1 & 7.83 & 3400 & 0.60 & 0.18 & 0.17 & 0.02 & 0.01 & $10^{-3}$ & $10^{-4}$ & 0.02 & $4\times 10^{-5}$\\
    13.1 & 7.82 & 3300 & 0.60 & 0.18 & 0.17 & 0.02 & 0.01 & $10^{-3}$ & $10^{-4}$ & 0.02 & $4\times 10^{-5}$ \\
    \hline
    \end{tabular}
    \label{tab:composition}
\end{table*}

\subsection{Host Metallicity}

We measured the emission line fluxes for a nearby \ion{H}{ii} region visible in the GMOS spectra, that is offset by $\sim$0.7 kpc from the location of \kyg. The spectrum has been corrected for the Milky Way extinction. The H$\beta$ line is located in the noisy blue region and we are thus unable to reliably measure the Balmer decrement from the ratio of H$\alpha$/H$\beta$. In order to estimate the metallicity, we adopted the \citet{2004MNRAS.348L..59P} calibration of the N2 method, which uses the log([\ion{N}{ii}]$\lambda$6583/H$\alpha$) ratio. It has the advantage of using two strong emission lines close to each other and thus it is not affected by host extinction. We obtain $12+{\rm log(O/H)}=8.68\pm0.04$ in N2 scale. 
In addition, we adopted the metallicity diagnostic of \cite{2016Ap&SS.361...61D}, that uses [\ion{N}{ii}]$\lambda$6583, H$\alpha$, [\ion{S}{ii}]$\lambda\lambda$6717,6731 lines, giving $12+{\rm log(O/H)}=8.85\pm0.10$. Assuming a solar oxygen abundance of $12+{\rm log(O/H)_{\odot}}=8.69$ \citep{2009ARA&A..47..481A}, we find the metallicity at this nearby \ion{H}{ii} region to be $\approx1.0-1.5\,Z_{\odot}$. This is comparable to the metallicity estimate for the Iax SN 2015H \citep{2017A&A...601A..62M}, and higher than the median explosion site metallicity for SNe Iax, $\sim 8.5$ dex \citep{2018MNRAS.473.1359L}.
While SNe Iax do not show a preference for sub-solar or super-solar metallicity environments in general \citep{2017A&A...601A..62M}, there is evidence for metal-poor environments for the faint Iax SNe 2008ha, 2010ae and 2019gsc \citep{2020ApJ...892L..24S}. Although \kyg\ does not seem to follow this trend, the number of events are small and it is not clear if there is a correlation between peak luminosity and metallicity.

\section{Volumetric rates of SNe Iax}\label{sec: rates}

In this section, we constrain the volumetric rate of SNe Iax in the local Universe and compare it to the general Ia rate within 100 Mpc. We specifically investigate the `faint' Iax subclass, where we define faint Iax events as those with peak absolute magnitude $M_r \gtrsim{} -16$, intermediate luminosity Iax events with $-17.5 \lesssim M_r \lesssim -16$ and bright Iax events with $M_r \lesssim -17.5$. There have been suggestions that these faint Iax SNe represent a physically distinct class of explosions and the volumetric rate is an important diagnostic tool for probing potential progenitor scenarios. 

\subsection{ATLAS Local Volume Survey}\label{subsec:ALVS}

We have constructed a 3.5 year sample of transients within a distance of 100 Mpc detected by the ATLAS survey (including transients discovered by other surveys), during 2017 September 21 (MJD 58017) and 2021 March 21 (MJD 59294). We will present a series of papers describing the full methodology and data sets and we briefly describe the ``ATLAS Local Volume Survey" here. There will be many objects in common with the ZTF Bright Transient Survey
\citep{2020ApJ...904...35P}, but our method of selecting objects with no magnitude constraint and association with host galaxies is different. \cite{2020ApJ...905...58D}  present the first results from a volume limited approach with ZTF.
We take the classifications as listed on the  
IAU Transient Name Server which tend to be  mostly from the ZTF broad effort and PESSTO \citep[][and its subsequent ePESSTO and ePESSTO+ incarnations]{2015A&A...579A..40S}. 
Comparison of results from ATLAS and ZTF, with different data, selection criteria and definitions will provide interesting discussion. 

On a typical night with two operating ATLAS units, we find $10-15$ real transient candidates (excluding variable stars, AGN, known movers etc.) in the  data stream. As described in  \citet{2020PASP..132h5002S}, after processing and filtering, these are registered on the Transient Name Server (TNS) as ``good" objects by a human scanner. A subset of these good objects are found to be associated with galaxies having a known redshift $z \lesssim 0.025$, corresponding to $D \lesssim 100$ Mpc. 
The radius of association with galaxies of known redshift is set at 50 kpc. When transients are found at large radial offsets, they are often background SNe, which are easily removed with close inspection of Pan-STARRS or SDSS images combined with spectra and the ATLAS light curves of the transients. The sample also includes confirmed SNe where the host galaxy redshift was unknown, but the SN redshift reported on the TNS, usually inferred from template matching tools such as SuperNova IDentification \citep[\textsc{snid;}][]{2007ApJ...666.1024B}, \textsc{gelato} \citep{2008A&A...488..383H}, \textsc{dash} \citep{2019ApJ...885...85M} or Superfit \citep{2005ApJ...634.1190H} was within our cutoff. In some cases the spectrum of the transient contained narrow emission lines from the host, providing a secure redshift. 
We adopt $H_0 = 70$ km s$^{-1}$ Mpc$^{-1}$ and apply a strict cut of $z \leq 0.024$ (corresponding to a co-moving distance of 102 Mpc). The adopted redshift $z$ for each transient is the associated host galaxy redshift if available, else the object redshift from the classification spectrum, as reported on the TNS.
The sample contains a total of 902 
transients, of which 134 
were without any spectroscopic data on TNS. 
The full volume-limited sample will be described in detail in a forthcoming paper (Srivastav et al., in prep.). 

The sample contains only six transients spectroscopically classified as SNe Iax. Of these, two are faint Iax  -- SNe 2019gsc \citep{2020ApJ...892L..24S} and 2020kyg. The four additional events that were classified as Iax in the sample are SNe 2019ovu \citep{2019TNSAN..90....1I}, 2019muj \citep{2019TNSCR1442....1H}, 2020sck \citep{2020TNSCR2685....1P} and 2020udy \citep{2020TNSCR2940....1N}. Polynomial fits to the ATLAS light curves suggests peak absolute magnitudes of $M_o = -16.35 \pm 0.10$ for SN 2019ovu, $M_o = -17.66 \pm 0.16$ for SN 2020sck, and $M_o = -17.81 \pm 0.15$ for SN 2020udy (corrected only for Galactic extinction). SN 2019muj is a well-observed event with peak $M_r = -16.35 \pm 0.08$ \citep{2021MNRAS.501.1078B}. SN 2020sck was studied in detail by \citet{2021arXiv211101226D}, who reported $M_R = -17.93 \pm 0.22$. SNe 2020sck and 2020udy are  bright Iax comparable to SN 2002cx (Figure~\ref{fig:PeakMag_dm15}), while SNe 2019ovu and 2019muj are of intermediate luminosity but above our threshold to be considered as part of the faint, low luminosity class of type Iax explosions. 
We note here that SN 2021fcg \citep{2021ApJ...921L...6K}, the faintest Iax event to date, was discovered within our 3.5 year window. However, it was not detected by ATLAS three times at $5\sigma$ significance on any single night, although ATLAS forced photometry does reveal $3\sigma$ detections during four epochs. 
Therefore we do not included it in our sample statistics.

\subsection{A search for further SNe Iax in the spectroscopically classified sample}

Since SN Iax spectra around maximum light are similar to SN Ia spectra, with the lower velocity being the primary distinction \citep{2013ApJ...767...57F}, 
there is a possibility that a few SNe Iax were misclassified as normal SNe Ia. In order to investigate this further, we analysed the $o$-band light curves for each of the 269 SNe Ia in our sample through ATLAS forced photometry \citep{2021TNSAN...7....1S}. 
A total of 73 were either old SNe Ia discovered well after peak with declining light curves, or did not have adequate coverage in ATLAS around peak for a reliable light curve fit. 
The remaining 196 light curves were fit with a low order polynomial function to estimate the time of $o$-band peak. For all these Ia events with at least one rising point on the ATLAS light curve and a minimum of 4 points around peak, the phase of the classification spectrum, with respect to the epoch of ATLAS $o$-band peak was determined. Since we are interested in measuring the velocity around maximum to discriminate between SNe Ia and Iax, only 158 events where the phase of the TNS classification spectrum was within $+15$d were selected for further analysis. Also, Si~{\sc ii} features in SN Ia spectra progressively weaken after maximum and blended features due to IGEs increasingly dominate, making a reliable estimate of \SiIIa\ velocity difficult beyond +15d. 

For the 158 SNe Ia that passed these cuts, we performed a Gaussian fit for the \SiIIa\ feature to estimate the photospheric velocity and pseudo-equivalent width (pEW). The fitting routine performs a continuum correction using points on the wings of the absorption feature selected interactively by the user. The errors on velocity and pEW were computed from multiple realisations of the Gaussian fit by adding small, random offsets to the selected continuum points. In general, the fitting procedure worked well, but failed for 1991T-like events that have shallow Si features, and a few old SNe Ia that managed to sneak through, despite our rising light curve criterion, since at later epochs this region of the spectrum is mostly dominated by blended features of IGEs. A reliable \SiIIa\ velocity could be estimated for a total of 113 SNe Ia in our sample.

\begin{figure*}
\includegraphics[width=0.8\textwidth]{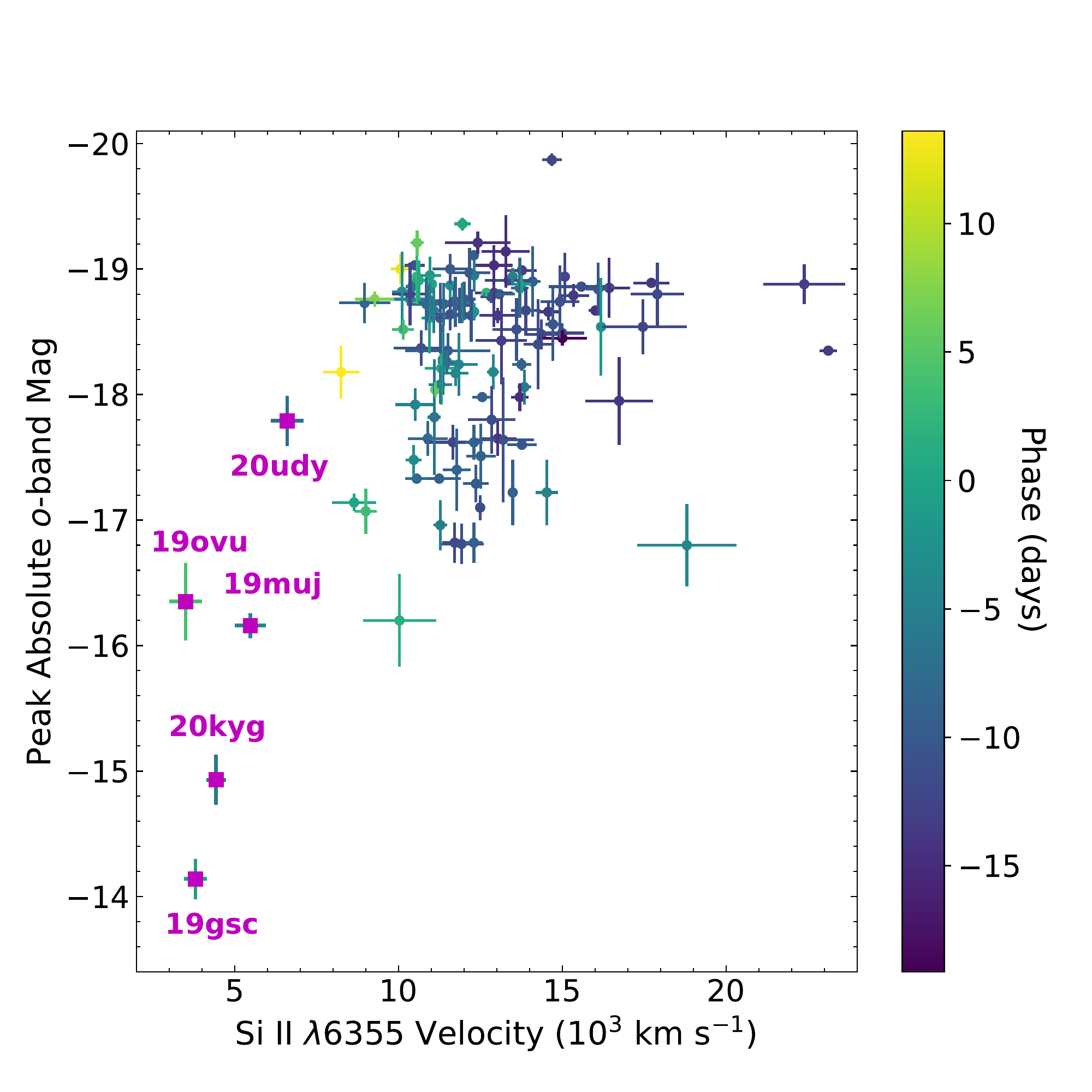}
\caption{Peak absolute $o$-band magnitude versus the \SiIIa\ velocity for 113 events in our sample classified as SNe Ia on the TNS. The colour map represents the phase of the classification spectrum with respect to the epoch of ATLAS $o$-band maximum. Also shown (magenta squares) are the SNe Iax in our sample, occupying a distinct region at low velocities. The colour of the error bars for the Iax in our sample represents the $o$-band phase of the spectrum used to deduce the \SiIIa\ line velocity.
}
\label{fig:AbsMagvsVel}
\end{figure*}

Figure~\ref{fig:AbsMagvsVel} shows the peak absolute ATLAS $o$-band magnitude versus the \SiIIa\ velocity for 113 SNe Ia with classification spectra within $+15$d. The phase of the classification spectrum relative to ATLAS $o$-band maximum is represented by the colour bar.
None of the SNe Ia in our sample show \SiIIa\ velocity $<8000$ \kms, the approximate boundary between the photospheric velocities of SNe Ia and Iax. Therefore, we don't find evidence for SNe Iax that were erroneously classified as SNe Ia in our sample. Also shown in Figure~\ref{fig:AbsMagvsVel} in magenta are the SNe Iax in our sample, except for SN 2020sck, as a reliable Si~{\sc ii} velocity could not be determined from the classification spectrum. This confirms, in a volume limited sample, that SNe Iax populate a distinct low velocity region and are easily distinguished from the bulk of the SN Ia population.

\subsection{A search for further faint SNe Iax in the unclassified sample}
\label{subsec:uncl}

In general, spectroscopic classification programs are likely to select brighter objects for classification, suggesting that the likelihood of intrinsically faint transients going unclassified is higher. Also, the short rise time ($\sim 10$ days) of faint SNe Iax means they are usually detected around maximum, limiting the window to obtain a good quality classification spectrum. There is a clear bias against 
spectroscopically classifying faint objects, and since this is a volume limited sample, intrinsically faint objects will 
preferentially lack classification. 

We have examined the light curves, distances, and host galaxies of all the unclassified transients in our sample to constrain how many are possibly faint Iax SNe in the ATLAS Local Volume Survey. There are 134 unclassified transients in the sample, as described in Section~\ref{subsec:ALVS}.
The extinction-corrected peak absolute magnitude of each unclassified event was determined using the relation: 

$$ M_{o,c}^{\rm peak} = m_{o,c}^{\rm peak} - A_{o,c} - \mu $$
Here, $m^{\rm peak}$ refers to the peak magnitude in the ATLAS orange or cyan band, $A$ is the Galactic extinction along the line of sight in the relevant band; $A_o \approx (A_r + A_i)/2$ and $A_c \approx (A_g + A_r)/2$, and $\mu$ is the distance modulus for the host galaxy. $\mu$ was calculated directly from the median redshift-independent distance for the host galaxy if available on NED. In case a redshift-independent distance was unavailable, $\mu$ was estimated from the redshift (assuming $H_0 = 70$ km s$^{-1}$ Mpc$^{-1}$). 
Applying a cut of $M^{\rm peak} \geq -16$ for faint Iax candidates brings down the number to 23 unclassified events. A majority (17) of these candidates show declining light curves (i.e. peak ATLAS mag $\equiv$ first ATLAS detection) and no recent history of non-detections prior to the first ATLAS detection, implying these are likely old SNe emerging from behind the Sun, and are thereby not considered as viable faint Iax candidates. 

The 6 remaining candidates are listed in Table~\ref{tab:uncl_candidates_tab}. AT 2018aes, has a very low luminosity of $M^{\rm peak}_{o} \approx -13.0 \pm 0.2$, and a red colour, $(c-o) \sim 0.3$, and has now been conclusively quantified as an Intermediate Luminosity Red Transient (ILRT)
by \citet{2021arXiv210805087C}. We used ATLAS forced photometry on the difference images at the positions of each of the 5 remaining candidates 
and determined the $o$-band peak magnitude and epoch using a polynomial fit (reported in  Table~\ref{tab:uncl_candidates_tab}). 
We discuss each of these below, and consider the two transients AT 2018atw and AT 2020jds as genuine faint Iax candidates. 

\subsubsection{AT 2019dgr}
 AT 2019dgr has  a slowly evolving light curve, quite unlike faint Iax. It  has a significant galactocentric offset of $R_{\rm g}\simeq 35$ kpc from its assumed host NGC 2987 ($D\simeq76$\,Mpc). This is more likely a background SN in a very faint host (undetected in both Pan-STARRS and SDSS) and unrelated to NGC 2987. The light curve shape suggests it is highly unlikely to be a faint SN Iax associated with NGC 2987. Thus we rule it out as a unclassified faint Iax candidate.
 
\subsubsection{AT 2019bds} 
\label{sec:19bds}
AT 2019bds is associated with a blue, irregular host galaxy in Pan-STARRS (PSO J148.5836$-$06.4883, 
$g_{\rm Kron}=18.3$). It has a redshift of 0.018 from the 2df Galaxy Redshift survey \citep{2003astro.ph..6581C}, and is a UV source (GALEXASC J095420.09-062916.7). At a projected offset of 1.9\arcsec (0.7 kpc),  the association seems secure. This implies it is more luminous ($M_o=-15.9$), and has a significantly broader light curve than known faint SNe Iax (Figure~\ref{fig:uncl_candidates}). Thus, we don't consider it to be a viable faint Iax candidate.

\subsubsection{AT 2018kae} AT2018kae is securely associated with ESO467-G027, being 2.4\,kpc offset from the core of this Sbc spiral galaxy. The transient has quite a peculiar asymmetric light curve, with a long rise of 20 days and a rapid decline of between 5-8 days to below $o\gtrsim20$. Since most transients are driven by a combination of $^{56}$Ni decay and diffusion timescales (leading to fast rise and exponential decay), an asymmetric light curve with a longer rise than decline can't easily be reproduced with a diffusion model for faint Iax SNe. We do not know the nature of this faint transient (perhaps a luminous blue variable outburst), but it is unlikely to be a faint Iax event.

\subsubsection{AT 2018atw} 
\label{sec:18atw}
AT2018atw is 4\,kpc offset from its likely host ESO 501-IG 092. This appears to be a merging or  interacting pair of galaxies with a visible tidal tail, and the 4\,kpc offset would not appear unusual. At this distance, the light curve resembles a faint Iax and this remains a viable candidate (Figure\,\ref{fig:uncl_candidates}).
The irregular host galaxy morphology is consistent with the typical late-type star-forming hosts of faint Iax, and SNe Iax in general.
   %
\subsubsection{AT 2020jds} AT 2020jds is offset by 3.7\,kpc from its secure host galaxy NGC 7535. At this distance of 50\,Mpc, it has a light curve shape and peak magnitude similar to the faint Iax sample. Although we don't have pre-discovery non-detection constraints for this transient, from the light curve it appears that it was discovered around maximum.
It was detected by Gaia the day before the ATLAS discovery, on MJD 58973 at $G=18.81$, and again on MJD 58988 at $G=19.89$, further indicating that we caught it on the rise to peak. We have also recovered detections in the Pan-STARRS NEO survey in the \zps\ (with Pan-STARRS2) and \wps\ (Pan-STARRS1) bands, which confirm the decline rate. 
The light curve indicates this is a viable unclassified faint Iax candidate. The spiral host is again consistent with Iax events, rather than the early-type host galaxies typically seen for the family of calcium-strong transients (CaSTs).

\begin{figure*}
	\includegraphics[width=0.8\linewidth]{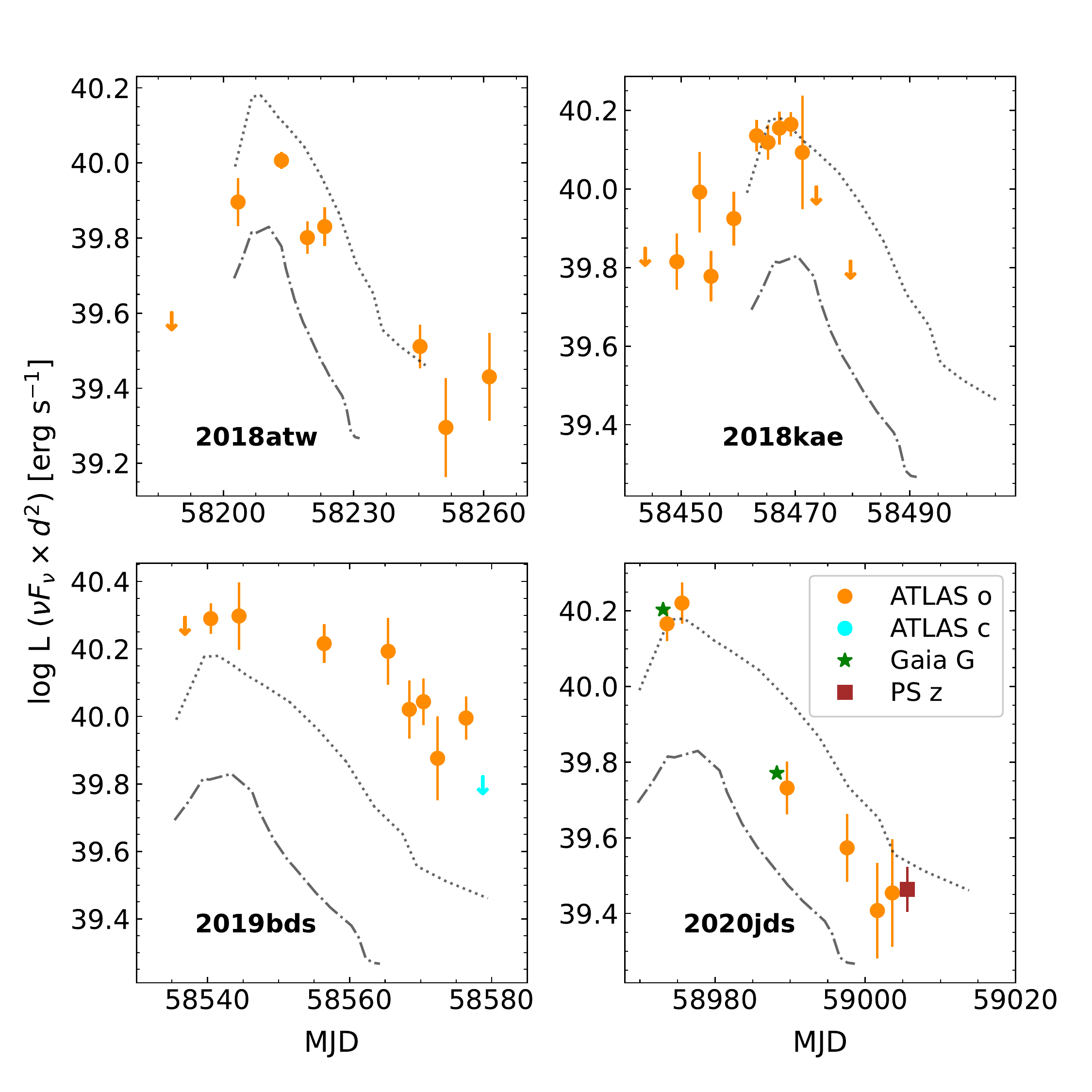}
    \caption{ATLAS forced photometry light curves of the unclassified faint transients ATs 2018atw, 2018kae, 2019bds and 2020jds. 
    For AT 2020jds, we also plot the available Gaia $G$-band and Pan-STARRS2 $z$-band photometry. For a direct comparison between different filters, we convert the magnitudes to a monochromatic luminosity as described in the text. Also shown are the light curves of \kyg\ (dotted lines) and SN 2008ha (dash-dotted lines), shifted along the time axis for comparison.
    }
    \label{fig:uncl_candidates}
\end{figure*}

After ruling out ATs 2018aes and 2019dgr, ATLAS forced photometry light curves of the remaining 4 unclassified candidates are shown in Figure~\ref{fig:uncl_candidates}. For AT 2020jds, we also plot the Gaia $G$-band and Pan-STARRS2 $z$-band photometry. In order to perform a direct comparison for photometry in different bands, we convert the magnitudes
to a monochromatic luminosity $L$ (in erg s$^{-1}$), using the relation
$$L = \nu_{\rm eff} \times F_{\nu,0} \times d^{2}$$
Here, $\nu_{\rm eff}$ is the effective filter frequency, $d$ is the assumed distance to the transient, and $F_{\nu,0} = F_{\nu} \times 10^{0.4A_{\nu}}$, is the extinction-corrected monochromatic flux density for a Galactic extinction $A_{\nu}$ in the relevant band.

~\\

\begin{table*}
\caption{ATLAS peak observed and absolute magnitudes of viable faint Iax candidates in the volume-limited sample that lacked a spectroscopic classification. The two confirmed faint Iax events SNe 2019gsc and 2020kyg are included for comparison. AT 2018aes is now a confirmed ILRT \citep{2021arXiv210805087C}.}
\begin{tabular}{cccccccc}\hline
TNS Name & Assumed Host Galaxy & Host Redshift & Distance & $\mu$ & Peak ATLAS Mag & Peak Abs. Mag$^\dagger$ & Comments\\
 & & & Mpc & mag & $o$-band & $o$-band \\
\hline
AT 2018aes & NGC 5300 & 0.003906 & 20.3$^*$ & 31.54 & $18.64 \pm 0.17$ & $-12.95$ & confirmed ILRT\\
AT 2019dgr & NGC 2987 & 0.012482 & 54.3$^*$ & 33.67 & $18.94 \pm 0.17$ & $-14.65$ & broad LC, background SN\\

AT 2018kae & ESO 467- G 027 & 0.017401 & 49.9$^*$ & 33.49 & $18.36 \pm 0.17$ & $-15.17$ & slow rise, rapid decline\\
AT 2019bds & 2dFGRS N034Z177 & 0.018400 & 75.8 & 34.4 & $18.62 \pm 0.17$ & $-15.90$ &  broad LC, faint Ibc?\\
\hline
AT 2018atw & ESO 501-IG 092 & 0.008613 & 37.2 & 32.85 & $18.35 \pm 0.12$ & $-14.62$ & candidate faint Iax \\
AT 2020jds & NGC 7535 & 0.015716 & 49.7$^*$ & 33.48 & $18.54 \pm 0.11$ & $-15.07$ & candidate faint Iax \\
\hline
SN 2019gsc & SBS 1436+529A & 0.011288 & 52.5 & 33.60 & $19.47 \pm 0.13$ & $-14.15$ & confirmed faint Iax \\
SN 2020kyg & NGC 5012 & 0.008736 & 43.4$^*$ & 33.17 & $18.23 \pm 0.05$ & $-14.97$ & confirmed faint Iax \\
\hline
\end{tabular}
\newline
\footnotesize{$^*$redshift independent distance on NED, otherwise distance based on redshift and $H_0=70$ \kms\,Mpc$^{-1}$} \\
\footnotesize{$^\dagger$ corrected for Galactic extinction}
\label{tab:uncl_candidates_tab}
\end{table*}


\subsection{Rate calculations}
\label{subsec:ratecalc}

The volumetric rate of a transient type is defined as:
\begin{equation}
R = \frac{N}{\epsilon VT}
\end{equation}
Here, $N$ is the number of events observed by ATLAS over time $T$ within a volume $V$. The efficiency factor $\epsilon$ represents the fraction of transients of a given type that are actually detected. $\epsilon$  accounts for the intrinsic luminosity of the transient type, sky coverage, variations in sensitivity or the $5\sigma$ limiting magnitude, breaks in observation due to weather patterns, the efficiency of the machine learning algorithm employed by the ATLAS transient server to identify real transients, and also the element of human error during the scanning process. The value of  $\epsilon$ should ideally also account for any dependence of detection on the galactocentric radius of the transient. The 
elevated sky background noise in higher surface brightness regions will reduce sensitivity, and image subtraction artefacts in the cores of galaxies may inhibit discovery of faint transients. 

We use the ATLAS efficiency simulation tool to estimate the recovery efficiency for faint Iax events \citep{ 2021PhDOwen}. For a given input light curve, the code generates a population of simulated light curves distributed randomly across a defined time window, redshift and sky coordinate range, and performs an assessment of recovery given the history of ATLAS observations. 
We have quality metrics for every ATLAS image 
(1,427,682 exposures)
including sky brightness,  image quality and most importantly the 5$\sigma$ limiting magnitude. We set the ATLAS survey declination limits at
$-50^{\circ} \leq \delta \leq 90^{\circ}$.  No right ascension limits are set since an event in solar conjunction may still be detectable well after peak. Our efficiency simulator accounts for this, and also accounts for Milky Way extinction, since all random positions chosen have foreground Milky Way extinction associated from the \citet{2011ApJ...737..103S} maps. The extinction in the ATLAS filters are estimated as $A_c = (A_g + A_r)/2$, and $A_o = (A_r + A_i)/2$. We do not yet account for additional internal host galaxy extinction, or the recovery efficiency as a function of galactocentric radius, although this will be implemented in future ATLAS Local Volume Survey rate papers (Srivastav et al. in prep).



The ATLAS transient server \citep{2020PASP..132h5002S} requires a minimum of three co-spatial, $5\sigma$ detections (out of the four dithered exposures obtained over the course of an hour) on a given night to consider a detection as a real astrophysical transient. However in practise, a transient candidate that has three detections just at the $5\sigma$ limit on only one night, is likely to be put on hold and not promoted to the TNS immediately by the human scanner. Detections on subsequent nights would tend to trigger the promotion and external registration on the TNS. To reproduce this real selection process, and effectively ensure that the simulated transient is detected on $\geq 3$ different nights, we enforce a minimum of 10 detections for a simulated transient to be considered as recovered. 

\begin{figure*}
	\includegraphics[width=0.7\linewidth]{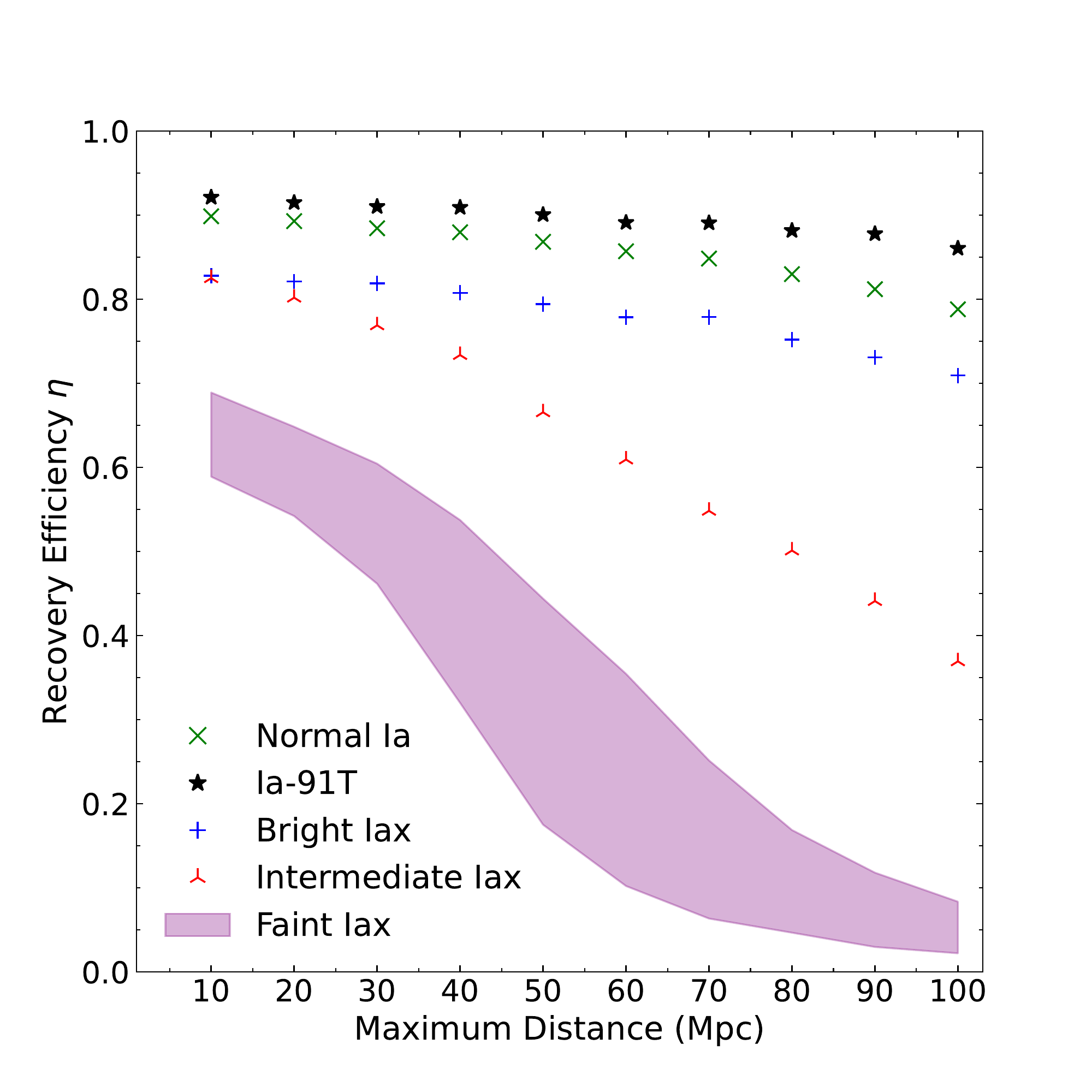}
    \caption{Simulated efficiency of recovery ($\eta$) in ATLAS for input light curves of 1991T-like Ia SN 2018cnw, normal Ia SN 2019ata, and bright, intermediate and faint Iax light curves as a function of distance. The bright Iax used for the simulation is SN 2020udy with peak luminosity $M_o = -17.81 \pm 0.15$, whereas the intermediate luminosity Iax is SN 2019muj with peak $M_o = -16.16 \pm 0.10$. The shaded region represents the range of $\eta$ for faint Iax events SNe 2020kyg and 2008ha, the latter being $\sim$ a magnitude fainter, and faster declining compared to \kyg.}
    \label{fig:efficiency}
\end{figure*}

The fraction of recovered transients to the total number simulated gives the efficiency of recovery ($\eta$). 
Of the 902 events in our 100 Mpc sample, 269 are spectroscopically confirmed SNe Ia. These include 9 SN 1991T-like and 9 SN 1991bg-like Ia  events. As expected, the efficiency of recovery in ATLAS is high for SN 1991T-like and normal SNe Ia light curves within 100 Mpc. The simulation suggests $\eta = 0.81$ for a normal type Ia, and $\eta = 0.86$ for a SN 1991T-like Ia, exploding at any time up to a distance of 100 Mpc (Figure~\ref{fig:efficiency}). 
However, the recovery efficiency for intermediate luminosity and faint Iax light curves is much lower, and falls off more steeply with increasing maximum distance, with $\eta = 0.35, 0.10$ for SNe 2020kyg, 2008ha respectively, for a volume corresponding to 60 Mpc.
\footnote{
The correct way of interpreting Figure\,\ref{fig:efficiency} is the recovery efficiency, $\eta$, {\em within the volume} enclosed by the distance (not the efficiency {\em at} that distance). } 

Correcting for the recovery efficiency and the geometrical factor for the area of sky available to the geographical site, the SN Ia rate within 100 Mpc from our sample (including {\rm only} those classified as type Ia SNe) is: 

$$R_{\rm Ia} (\rm classified) \approx 2.41 \pm 0.25 \times 10^{-5} \times h_{70}^{3}\, \mathrm{Mpc^{-3}\,yr^{-1}}$$ 

This is consistent with the 
volumetric Ia rate from PTF, $R_{\rm Ia}=2.43^{+0.33}_{-0.19} \pm 0.29 \times 10^{-5}$ Mpc$^{-3}$ yr$^{-1}$ \citep{2019MNRAS.486.2308F}, and the  ZTF Bright Transient Survey  \citep{2020ApJ...904...35P}, who found
$R_{\rm Ia}=2.35\pm 0.24 \times 10^{-5}$ Mpc$^{-3}$ yr$^{-1}$. 
Here, $h_{70} = H_0/70$ is the correction factor for the adopted value of the Hubble's constant. The systematic error on the SN Ia rate quoted above is derived from the error on the 100 Mpc volume $V$ (corresponding to $z = 0.024 \pm 0.002$), and the recovery efficiency $\eta = 0.81 \pm 0.04$.

A sizeable number of transients that are almost certainly SNe lack a spectroscopic classification. Most of these have detections over mutliple nights, clear association with a host galaxy within 100 Mpc, and a light curve that is consistent with being a SN. Many were unclassified as they had incomplete light curves, 
often in the tail phase. If we assume that the 134 unclassified candidates have the same relative 
fractions of different SN types as the classified sample, this would suggest an additional $\sim 47$ SNe Ia, implying
$$R_{\rm Ia} (\rm total) \approx 2.83 \pm 0.29 \times 10^{-5} \times h_{70}^{3}\,\mathrm{Mpc^{-3}\,yr^{-1}}$$


This is consistent with the \citet{2011MNRAS.412.1441L} rate, $R_{\rm Ia}=3.0 \pm 0.1 \times 10^{-5}$ Mpc$^{-3}$ yr$^{-1}$, although it relies on the assumption that the unclassified sample contains similar ratios of SN types as the classified sample. For those discovered well after peak, this is likely a reasonable assumption, but further work is required, such as simulating the rates and confirming the number of late phase detections expected.


Given the typical $5\sigma$ limiting magnitudes of $19-19.5$, ATLAS would not detect faint SNe Iax like 2019gsc and 2008ha beyond $\sim 50-60$ Mpc. We thus limit our rate calculations for faint Iax to a volume corresponding to 60 Mpc. Figure\,\ref{fig:efficiency}
shows the efficiency of recovery diminishes rapidly beyond $50-60$ Mpc. The luminosity function for faint Iax events (shaded region in Figure~\ref{fig:efficiency}) is represented by \kyg\ (bright end) and SN 2008ha (faint end). For a distance of up to 60 Mpc, the recovery efficiencies for 2020kyg-like and 2008ha-like light curves are $\eta = 0.36, 0.10$, respectively. We thus assume a combined recovery efficiency of $\eta = 0.23 \pm 0.13$ for faint Iax. 
This effectively assumes a flat luminosity function with a sharp cut-off at each end. Given we have only a few well-observed events, such a simplistic function appears justified. Within 60 Mpc, the faint Iax rates, given two confirmed events is: 


$$R_{\rm fIax} (N=2) \approx 2.92_{-1.89}^{+3.86} \pm 1.06 \times 10^{-6} \times h_{70}^{3} \, \mathrm{Mpc^{-3}\, yr^{-1}}$$ 


The asymmetric statistical uncertainties on the rate represent $1\sigma$ Gaussian errors computed from single sided upper and lower limits for small number statistics \citep{1986ApJ...303..336G}, whereas the systematic uncertainties include the error on the recovery efficiency $\eta$.
It is quite plausible, if not probable, that we are spectroscopically incomplete for faint type Iax, due 
a bias against spectroscopically classifying faint events. We have demonstrated that the two unclassified transients ATs 2020jds and 2018atw are plausible faint Iax candidates (Sections\,\ref{sec:19bds}, \ref{sec:18atw} and Figure\,\ref{fig:uncl_candidates}). 
Hence if these are also members of the subclass, the rate would be: 
$$R_{\rm fIax} (N=4) \approx 5.85_{-2.80}^{+4.62} \pm 2.11 \times 10^{-6} \times h_{70}^{3} \, \mathrm{Mpc^{-3}\, yr^{-1}}$$ 

For the intermediate and bright Iax categories, it is not straightforward to estimate the number of potential unclassified candidates from their light curves, since their luminosity range overlaps with SNe Ib/c and sub-luminous SNe Ia. We compute a rough estimate for the number of unclassifed candidates by scaling their recovery efficiency with respect to SNe Ia. For example, potential number of unclassified bright SNe Iax is estimated as follows:

$$N_{\rm bIax}^{\rm unclassified} \sim \frac{N_{\rm bIax}^{\rm classified}}{N_{\rm Ia}^{\rm classified}} \times N_{\rm Ia}^{\rm unclassified} \times \frac{\eta_{\rm Ia}}{\eta_{\rm bIax}}$$
Here, $N_{\rm Ia}^{\rm unclassified} \approx 47$, as assumed above.

The volumetric rates for SNe Ia and the Iax categories are summarised in Table~\ref{tab:rates}.
The potential number of events from the unclassified sample estimated above is included in the systematic uncertainty for the rates of bright and intermediate SNe Iax.
Since we have demonstrated robustly that there are likely to be very few intermediate and bright type Iax in the classified sample which have been missed (Figure\,\ref{fig:AbsMagvsVel}), the true volumetric rate of these types must 
be low. In fact the rate of SNe Iax is dominated by the faint members of this spectroscopic class. The bright SNe Iax (see Figure\,\ref{fig:PeakMag_dm15}, 
$M_r \lesssim -17.5$) make up $0.9^{+1.1}_{-0.5}\%$ of the SN Ia rate (a 2$\sigma$ upper limit of $\sim3$\%), 
and combined with the intermediate luminosity objects 
(see Table\,\ref{tab:rates}), the relative rate of type Iax with $M_r \lesssim -16$ is $3^{+5}_{-2}$\%.


We estimate that the faint Iax rate is much higher at $12^{+14}_{-8}\%$ of the SN Ia rate and hence they dominate the volumetric rate for the inhomogeneous type Iax class. The very recent discovery of SN 2021fcg \citep{2021ApJ...921L...6K} indicates that even fainter SNe Iax could exist, further enhancing the rates. 
Our simulations suggest that the recovery efficiency of a 2021fcg-like light curve is only $\eta = 0.02$ within 60 Mpc. This yields a combined $\eta = 0.19 \pm 0.17$, enhancing the faint Iax rates to $15^{+18}_{-10}\%$ of the SN Ia rate.
The large uncertainty is not surprising if such extremely faint objects contribute significantly to the luminosity function, since SN~2021fcg is so faint it was not independently detected in our ATLAS data. 
In addition, the form of the luminosity function between absolute peak magnitudes of $-15$
and $-12$ (see Figure\,\ref{fig:PeakMag_dm15}) is unconstrained.





\renewcommand{\arraystretch}{1.5}
\begin{table*}
\caption{Summary of the rate calculations for SNe Ia and SNe Iax (bright, intermediate and faint) from the 100 Mpc ATLAS Local Volume Survey. The distance column represents the maximum distance out to which simulated transients were placed for calculating recovery efficiency. The volumetric rate computed for the spectroscopically classified events represents a lower limit. A total rate is also estimated (last column) from considering plausible candidates in the unclassified sample.}
\begin{tabular}{ |c|c|c|c|c|c|c|c| }\hline
SN & Distance & \multicolumn{3}{|c|}{Number of events} & Recovery  & \multicolumn{2}{|c|}{Rate ($h_{70}^{3}$ Mpc$^{-3}$ yr$^{-1}$)} \\
Type & Mpc & Classified & Unclassified & Total & Efficiency & Classified & Total \\
 \hline
 Ia & 100 & 269 & 47 & 316 & $0.81\pm 0.04$ &  $2.41 \pm 0.25 \times 10^{-5}$ & $2.83 \pm 0.29 \times 10^{-5}$ \\ 
 Bright Iax & 100 & 2 & 0.4$^{*}$ & 2.4 & $0.71\pm 0.05$ & $2.05^{+2.70}_{-1.32} \pm 0.25 \times 10^{-7}$ & $2.45^{+2.95}_{-1.39} \pm 0.27 \times 10^{-7}$\\
 Intermediate Iax & 100 & 2 & 0.8$^{*}$ & 2.8 & $0.37\pm 0.08$ & $3.93^{+5.18}_{-2.54} \pm 1.16 \times 10^{-7}$ & $5.43^{+6.19}_{-2.74} \pm 1.44 \times 10^{-7}$ \\
Faint Iax & 60 & 2 & 2 & 4 & $0.23\pm 0.13$ & $2.92^{+3.86}_{-1.89} \pm 1.06 \times 10^{-6}$ & $5.85^{+4.62}_{-2.80} \pm 2.11 \times 10^{-6}$ \\
\hline
\end{tabular}
\label{tab:rates}
\newline
$^{*}$Computed by scaling the recovery efficiency with respect to that of SNe Ia, as described in Section~\ref{subsec:ratecalc}.
\end{table*}

\section{Discussion and Conclusions}

We have presented the results of our multi-wavelength follow-up campaign for the faint Iax \kyg. We define faint SNe Iax as events with peak luminosity $M_r \gtrsim -16$. This subclass only consists of a handful of other well-observed events such as SNe 2008ha \citep{2009AJ....138..376F,2009Natur.459..674V}, 2010ae \citep{2014A&A...561A.146S}, 2019gsc \citep{2020ApJ...892L..24S,2020MNRAS.496.1132T} and 2021fcg \citep{2021ApJ...921L...6K}. With a peak luminosity of $-15 \lesssim M_r \lesssim -13$, these events represent the faint extremity of SN explosions. 

\begin{figure*}
\includegraphics[width=0.8\linewidth]{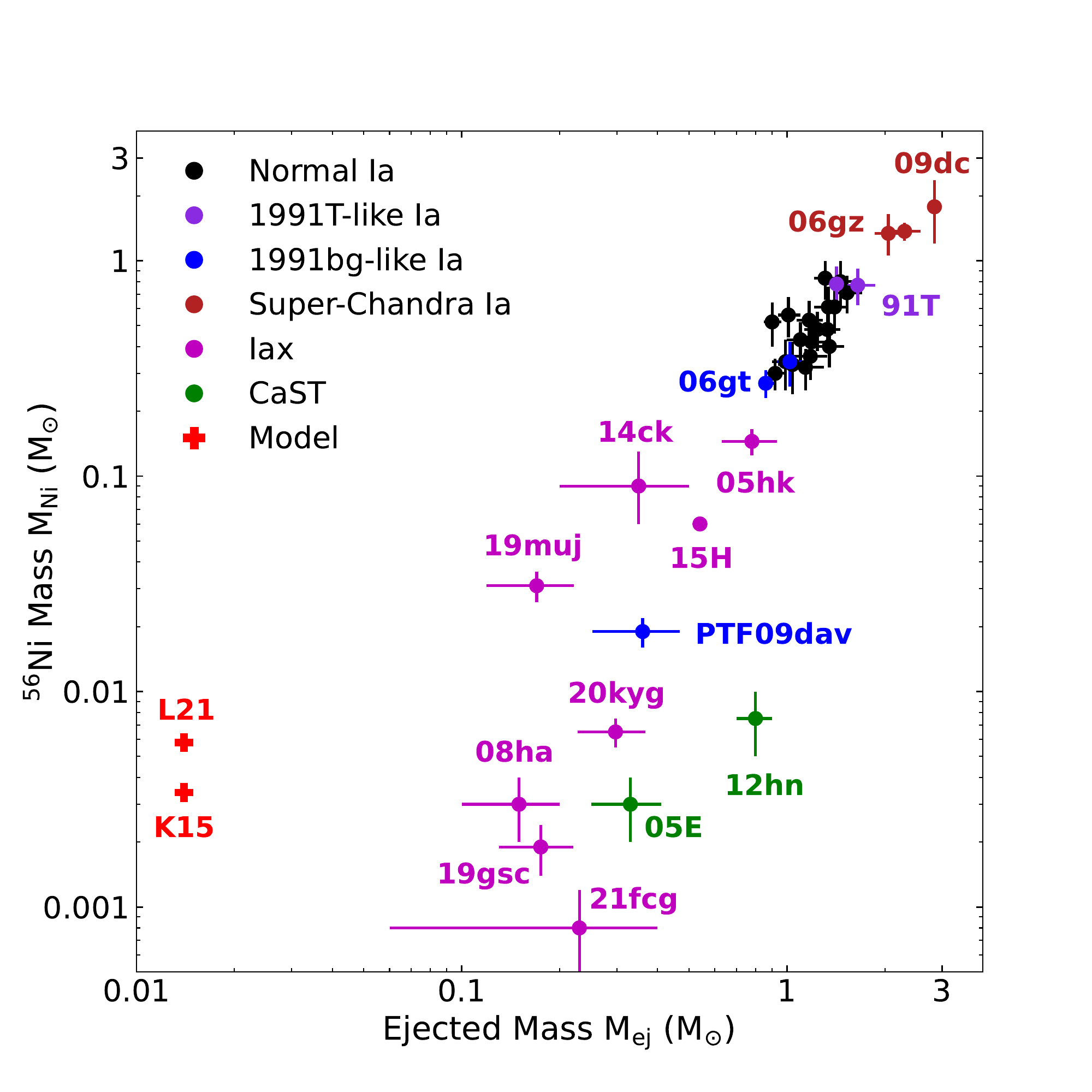}
\caption{The $M_{\rm Ni}-M_{\rm ej}$ parameter space for SNe Ia and CaSTs, adapted from \citet[][and references therein]{2014ApJ...786..134M}. Additional references: Normal Ia \citep{2014MNRAS.445.2535S}, SNe 2006gt, 2006gz, 2008ec \citep{2019MNRAS.483..628S}, SN 2014ck \citep{2016MNRAS.459.1018T}, SN 2015H \citep{2016A&A...589A..89M}, SN 2019gsc \citep{2020ApJ...892L..24S}, SN 2019muj \citep{2021MNRAS.501.1078B}, SN 2020kyg (this work) and SN 2021fcg \citep{2021ApJ...921L...6K}. The \textcolor{red}{\ding{58}} symbols represent two of the faintest deflagration simulations in the literature - the hybrid CONe WD model of \citet[][or K15]{2015MNRAS.450.3045K} and the \texttt{r114\textunderscore d6.0\textunderscore Z} model of \citet[][or L21]{2021arXiv210902926L}.
} 
\label{fig:MniMej}
\end{figure*}

Analytical modeling of the quasi-bolometric light curves of these events (Figure~\ref{fig:bolometric}) suggests $^{56}$Ni masses in the range of $(1-7) \times 10^{-3}$ \msol, roughly two orders of magnitude lower than that for normal SNe Ia. The inferred ejecta masses, in the range $0.2-0.4$ \msol, are not as extreme, implying very low $M_{\rm Ni}/M_{\rm ej}$ ratios. Figure~\ref{fig:MniMej} shows $M_{\rm Ni}$ versus $M_{\rm ej}$ for different classes of thermonuclear SNe, adapted from \citet[][Figure 15]{2014ApJ...786..134M}.
We also plot two points from the 3D deflagration simulations that have been calculated to try and model the full luminosity range of SNe Iax \citep{2015MNRAS.450.3045K,2021arXiv210902926L}.
The models can produce $M_{\rm Ni}$ yields that are broadly consistent with the low inferred values for faint SNe Iax, but the predicted $M_{\rm ej}$ is too low when compared to the 
observationally inferred values. 
The inferred rise times are $\sim 10$ days for these events. For \kyg\ and SN 2019gsc, these inferred rise times are consistent with the constraints from pre-discovery non-detections in ATLAS and ZTF before the light curve rises. 

Spectroscopically, these faint events display lower expansion velocity in general, but otherwise resemble more luminous SNe Iax albeit with more rapid evolution. We don't see evidence for interaction with circumstellar material (CSM) in the spectra, and the quasi-bolometric light curves are adequately fit with a purely radioactive source of energy. The \SiIIa\ velocity evolution (Figure~\ref{fig:SiIIvel}) displays a trend with fainter SNe Iax displaying lower velocities. Indeed, this correlation has been investigated in the literature \citep[eg.][]{2010ApJ...720..704M}. However, more luminous members of the subclass, such as SNe 2009ku \citep{2011ApJ...731L..11N} and 2014ck \citep{2016MNRAS.459.1018T} displayed very low velocities $\sim 3000$ \kms\ comparable to the faint members, defying a single parameter description \citep{2017A&A...601A..62M,2018MNRAS.480.3609B}. We use \tardis\ to model the early photospheric spectra of \kyg\ between $-3$ and $+4$d (Figure~\ref{fig:tardisfit}) with a simple uniform abundance model dominated by carbon, oxygen, neon and other IMEs, similar to that used for SN 2019gsc \citep{2020ApJ...892L..24S}. The same model when computed at +16d produces a poor fit to the observed spectrum, that is already dominated by IGEs. This could suggest a highly mixed ejecta with significant radioactive material in the outer layers, that would limit the applicability of \tardis\ at this later epoch.

The overall observed photometric and spectroscopic properties of SNe Iax, including faint Iax, suggest these are related to SNe Ia and thus share a thermonuclear origin \citep{2017hsn..book..375J}. 3D simulations have shown that weak deflagrations can be unsuccessful in fully unbinding the WD progenitor, leaving a compact bound remnant behind \citep{2016A&A...589A..38B}. A promising pathway for faint SNe Iax is a weak deflagration involving a hybrid CONe WD \citep{2013ApJ...772...37D} instead of a CO WD progenitor, and 3D simulations of such a near-\mch\ WD by \citet{2015MNRAS.450.3045K} show this scenario can account for extremely low $^{56}$Ni yields. This model produces a very low ejecta mass, implying the bound remnant is near-\mch. Binary population synthesis calculations have shown that hybrid CONe WDs accreting from a helium-rich donor have short delay times, as low as 30 Myr \citep{2014ApJ...789L..45M,2014ApJ...794L..28W,2015A&A...574A..12L,2015MNRAS.450.3045K}. This is consistent with the results of \citet{2018MNRAS.473.1359L}, who found evidence for young stellar populations in the environments of most SNe Iax in a sizeable sample. The six Iax events in our sample are hosted by late-type, spiral or irregular galaxies.
Additionally, these hybrid CONe WDs could be more massive at formation, up to $\sim 1.3$ \msol\ \citep{2014MNRAS.440.1274C}, thus requiring even shorter delay times. The blue source in pre-explosion images coincident with the location of SN Iax 2012Z was interpreted as the helium star companion of the WD progenitor \citep{2014ApJ...786..134M}. This source persists in HST images obtained 1400 days after explosion \citep{2021arXiv210604602M}. The late time flux was found to be higher than the pre-explosion flux by a factor of two, suggesting that the bound remnant is likely contributing to this excess. We favor the hybrid CONe WD + He star scenario for explaining faint Iax events.

If the progenitors are indeed WD + He stars, the interaction of SN ejecta with the companion would be expected to strip off helium-rich material from the donor \citep[eg.][]{2019ApJ...887...68B}, suggesting there could be signatures of helium in late-time spectra of SNe Iax. Attempts to find helium in late-time spectra of SNe Iax have been unsuccessful so far \citep{2019MNRAS.487.2538J,2019A&A...622A.102M}. However, 3D simulations by \citet{2020ApJ...898...12Z} managed to strip off only $4\times 10^{-3}$ \msol\ of helium from the companion, consistent with the upper limits from observations. Thus, the lack of observed helium in the spectra does not necessarily rule out its presence in the ejecta.

In order to constrain the volumetric rates of SNe Iax and in particular, faint Iax, we construct a homogeneous sample of 902 transients within a distance of $100$ Mpc observed by ATLAS. The rates of SNe Iax are known to be dominated by lower luminosity events \citep{2017ApJ...837..121G}. Our 100 Mpc sample contains only six Iax events, of which two are faint Iax events, two intermediate luminosity and two luminous Iax events. Our derived volumetric rate for SNe Ia is $2.83 \pm 0.29 \times 10^{-5}$ Mpc$^{-3}$ yr$^{-1}$, consistent with the Lick Observatory Supernova Search (LOSS) rates in \citet{2011MNRAS.412.1441L}. 
The derived volumetric rate for faint SNe Iax (within 60 Mpc) is $2.92_{-1.89}^{+3.86} \pm 1.06 \times 10^{-6}$ Mpc$^{-3}$ yr$^{-1}$, accounting for $12^{+14}_{-8}\%$ of the SN Ia rate. These rates are consistent with binary population synthesis calculations for CONe WD + He star systems by \citet{2014ApJ...794L..28W}, who proposed this pathway could account for $1-18\%$ of the Ia rate. The overall Iax rate is $15^{+17}_{-9}\%$ of the SN Ia rate, clearly dominated by faint SNe Iax. Luminous Iax events like SNe 2002cx and 2005hk are relatively rare, accounting for $0.9^{+1.1}_{-0.5}\%$ of the Ia rate.

The volumetric rates suggest there could be $40-100$ SN Ia remnants in the Milky Way \citep{2021ApJ...908...31Z}. Given our constraints on the relative rates of SNe Iax, this would imply $3-25$ Iax SNRs in our Galaxy. In the CONe WD + He star channel for SNe Iax, in addition to the partially burnt bound remnant or primary remnant (PR), the kicked companion or donor remnant (DR), that may have evolved to the WD stage, is also expected to survive \citep{2019MNRAS.489..420R}. In principle, both PRs and DRs could be detectable in the Galaxy, although SN Ia remnants may be more difficult to detect compared to CCSN remnants \citep{2017MNRAS.464.2326S}. Of late, a handful of WDs with peculiar velocities, unusual mass/radii and composition with evidence of enrichment in IMEs have been discovered \citep[eg.][]{2017Sci...357..680V,2018MNRAS.479L..96R,2018ApJ...865...15S}. These hyper velocity WDs have been interpreted as the kicked DRs from a Iax explosion. Using X-ray spectroscopy, \citet{2021ApJ...908...31Z} found high Mn/Fe and Ni/Fe ratios in the SNR Sgr A East, interpreting it as the surviving PR from a Iax event involving a CO WD. Thus, alongside deep pre and post-explosion observations of nearby extragalactic SNe Iax, searching and characterising such peculiar hyper velocity WD systems in the Galaxy is also a promising avenue for piecing together the progenitor puzzle. 

\section*{Acknowledgements}

SS, SJS and SAS acknowledge funding from STFC Grants  ST/P000312/1 and ST/T000198/1.
CA and JH were supported by a VILLUM FONDEN Investigator grant to JH (project number 16599). KA was supported by Australian Research Council Centre of Excellence for All Sky Astrophysics in 3 Dimensions (ASTRO 3D), through project number CE170100013. GP is supported by ANID - Millennium Science Initiative - ICN12\_009. TWC acknowledges the EU Funding under Marie Sk\l{}odowska-Curie grant H2020-MSCA-IF-2018-842471.


Pan-STARRS is a project of the Institute for Astronomy of the University of Hawaii, and is supported by the NASA SSO Near Earth Observation Program under grants 80NSSC18K0971, NNX14AM74G, NNX12AR65G, NNX13AQ47G, NNX08AR22G, 80NSSC21K1572 and by the State of Hawaii.

This work has made use of data from the Asteroid Terrestrial-impact Last Alert System (ATLAS) project. The Asteroid Terrestrial-impact Last Alert System (ATLAS) project is primarily funded to search for near earth asteroids through NASA grants NN12AR55G, 80NSSC18K0284, and 80NSSC18K1575; byproducts of the NEO search include images and catalogs from the survey area. This work was partially funded by Kepler/K2 grant J1944/80NSSC19K0112 and HST GO-15889, and STFC grants ST/T000198/1 and ST/S006109/1. The ATLAS science products have been made possible through the contributions of the University of Hawaii Institute for Astronomy, the Queen’s University Belfast, the Space Telescope Science Institute, the South African Astronomical Observatory, and The Millennium Institute of Astrophysics (MAS), Chile.

The data presented here were obtained in part with ALFOSC under programme 61-022 (PI Angus), which is provided by the Instituto de Astrofisica de Andalucia (IAA) under a joint agreement with the University of Copenhagen and NOT.

Based in part on observations obtained with MegaPrime/MegaCam, a joint project of CFHT and CEA/DAPNIA, at the Canada-France-Hawaii Telescope (CFHT) which is operated by the National Research Council (NRC) of Canada, the Institut National des Science de l'Univers of the Centre National de la Recherche Scientifique (CNRS) of France, and the University of Hawaii. The observations at the Canada-France-Hawaii Telescope were performed with care and respect from the summit of Maunakea which is a significant cultural and historic site.

This work was enabled in part by observations made from Gemini North telescope and UKIRT, located within the Maunakea Science Reserve and adjacent to the summit of Maunakea. We are grateful for the privilege of observing the Universe from a place that is unique in both its astronomical quality and its cultural significance. UKIRT is owned by the University of Hawaii (UH) and operated by the UH Institute for Astronomy. When the data reported here were obtained, the operations were enabled through the cooperation of the East Asian Observatory (EAO).

The Liverpool Telescope is operated on the island of La Palma by Liverpool John Moores University in the Spanish Observatorio del Roque de los Muchachos of the Instituto de Astrofisica de Canarias with financial support from the UK Science and Technology Facilities Council.

Based in part on observations obtained at the international Gemini Observatory, a program of NSF’s NOIRLab, which is managed by the Association of Universities for Research in Astronomy (AURA) under a cooperative agreement with the National Science Foundation. on behalf of the Gemini Observatory partnership: the National Science Foundation (United States), National Research Council (Canada), Agencia Nacional de Investigaci\'{o}n y Desarrollo (Chile), Ministerio de Ciencia, Tecnolog\'{i}a e Innovaci\'{o}n (Argentina), Minist\'{e}rio da Ci\^{e}ncia, Tecnologia, Inova\c{c}\~{o}es e Comunica\c{c}\~{o}es (Brazil), and Korea Astronomy and Space Science Institute (Republic of Korea).

This work made use of Astropy,\footnote{http://www.astropy.org} a community-developed core Python package for Astronomy \citep{2013A&A...558A..33A,2018AJ....156..123A} and the WISeREP archive - \url{https://www.wiserep.org/} \citep{2012PASP..124..668Y}.

\section*{Data Availability}

The spectroscopic data for \kyg\ underlying this article will be available at the WISeREP archive, \url{https://www.wiserep.org/}.



\bibliographystyle{mnras}
\bibliography{references} 




\appendix



\bsp	
\label{lastpage}
\end{document}